\documentclass[10pt,twoside]{article}
\usepackage{amsfonts}
\usepackage{amssymb}
\usepackage{amsmath}
\usepackage{pifont}
\usepackage{graphicx}
\usepackage[english]{babel}

\numberwithin{equation}{section}

\begin{document}

\title{$SU(3)_{c}\otimes SU(3)_{L}\otimes U(1)_{X}$ models for $\beta $
arbitrary and families with mirror fermions}
\author{Rodolfo A. Diaz\thanks{%
radiazs@unal.edu.co}, R. Martinez\thanks{%
remartinezm@unal.edu.co}, F. Ochoa\thanks{%
faochoap@unal.edu.co} \\
Universidad Nacional de Colombia, \\
Departamento de F\'{\i}sica. Bogot\'{a}, Colombia.}
\date{}
\maketitle

\vspace{-5mm}

\begin{abstract}
A detailed and general study of the fermionic structure of the 331 models
with $\beta $ arbitrary is carried out based on the criterion of
cancellation of anomalies. We consider models with an arbitrary number of
lepton and quark generations, but requiring to associate only one lepton and
one quark $SU\left( 3\right) _{L}$ multiplet for each generation, and at
most one right-handed singlet per each left-handed fermion. We see that the
number of quark left-handed multiplets must be three times the number of
leptonic left-handed multiplets. Furthermore, we consider a model with four
families and $\beta =-1/\sqrt{3}$ where the additional family corresponds to
a mirror fermion of the third generation of the Standard Model. We also show
how to generate ansatzs about the mass matrices of the fermions according to
the phenomenology. In particular, it is possible to get a natural fit for
the neutrino hierarchical masses and mixing angles. Moreover, by means of
the mixing between the third quark family and its mirror fermion, a possible
solution for the $A_{FB}^{b}$ discrepancy is obtained.


\end{abstract}

\vspace{-6mm}

\section{Introduction}

A very commom alternative to solve some of the problems of the standard
model (SM) consists of enlarging the group of gauge symmetry, where the
larger group embeds properly the SM. For instance, the $SU(5)$\ grand
unification model of Georgi and Glashow \cite{seven} can unify the
interactions and predicts the electric charge quantization; while the group $%
E_{6}$ can also unifies the interactions and might explain the masses of the
neutrinos \cite{nine}. Nevertheless, such models cannot explain the origin
of the fermion families. Some models with larger symmetries address this
problem \cite{familia}. A very interesting alternative to explain the origin
of generations comes from the cancellation of chiral anomalies \cite%
{anomalias}. In particular, the models with gauge symmetry $SU(3)_{c}\otimes
SU(3)_{L}\otimes U(1)_{X},$ also called 331 models, arise as a possible
solution to this puzzle, since some of such models require the three
families in order to cancel chiral anomalies completely. An additional
motivation to study these kind of models comes from the fact that they can
also predict the charge quantization for a three family model even when
neutrino masses are added \cite{Pires}. Finally, supersymmetric versions of
this gauge theory have also been studied \cite{PLBus}.

Despite the 331 models could formally provide an explanation for the number
of families, they cannot explain many aspects that the SM cannot explain
either, it suggests the presence of new physics. In the current versions of
the model it is not possible to explain the mass hierarchy and mixing of the
fermions. On the other hand, the model is purely left-handed, so that it
cannot account about parity breaking. Another point of interest to study in
the models is the CP violation, particularly the strong CP violation which
might allow us to understand the values for the electric dipole moment of
the neutron and electron.

Although cancellation of anomalies leads to some conditions \cite{fourteen},
such criterion alone still permits an infinite number of 331 models. In
these models, the electric charge is defined in general as a linear
combination of the diagonal generators of the group 
\begin{equation}
Q=T_{3}+\beta T_{8}+XI,  \label{charge}
\end{equation}%
As it has been extensively studied in the literature \cite{fourteen, ten, eleven}, the value of the $\beta $ parameter determines the fermion
assignment and more specifically, the electric charges of the exotic
spectrum. Hence, it is customary to use this quantum number to classify the
different 331 models. If we want to avoid exotic charges we are led to only
two different models i.e. $\beta =\pm 1/\sqrt{3}$ \cite{fourteen, twelve}.

In the analysis for $\beta $ arbitrary based on the cancellation of
anomalies, we find many possible structures that contain the SM at low
energies. In the model with two leptonic left-handed multiplets ($N=2$), we
get a one family model in which one of the multiplets correspond to the
mirror fermions (MF)\ of the other, i.e., the quarks and leptons form vector
representations with respect to $SU\left( 3\right) _{L}$ for each family.
Two additional copies are necessary in order to obtain the SM at low
energies.

The structure for $N=4$ families and $\beta =-1/\sqrt{3}$, where three of
them refer to the generations at low energies and the other is a mirror
family, is a vector-like model that has two multiplets in the $\mathbf{3}$
representation and two multiplets in the $\mathbf{3}^{\ast }$ representation
in both the quark and lepton sectors. This extension of the 331 model is not
reduced to the known models with $\beta =-\sqrt{3},-1/\sqrt{3}$ \cite{ten,
twelve}, because in such models the leptons are in three 3-dimensional
multiplets. From the phenomenological point of view at low energies, the
difference would be in generating ansatz for the mass matrices in the lepton
and quark sectors. Models with vector-like multiplets are necessary to
explain the family hierarchy. Moreover, it is observed that the
neutrinos do not exhibit a strong family hierarchy pattern as it happens
with the other fermions; the mixing angles for the neutrinos $\theta _{atm}$
and $\theta _{sun}$ are not small; besides, the quotient $\left( \delta
m_{sun}^{2}/\delta m_{atm}^{2}\right) $ is of the order of $0.02-0.03$, these facts
suggest to modify the see-saw mechanism in order to cancel the hierarchy in
the mass generation for the neutrinos, such modifications are usually
implemented by introducing vector-like fermion multiplets \cite{Valle}.

On the other hand, the deviation of the $b-$quark asymmetry $A_{b}$ from the
value predicted by SM (of the order of $3\sigma $), suggests a modification
in the right-handed couplings of $Z_{\mu }$ with the $b-$quark, by means of
particles that are not completely decoupled at low energies. An alternative
is the inclusion of MF because they acquire masses slightly greater than the
electroweak scale since their masses are generated when $SU\left( 2\right)
_{L}\otimes U\left( 1\right) _{Y}$ is broken \cite{Minkowski}. Further, a
model with MF couples with right-handed chirality to the electroweak gauge
fields. Hence, these couplings might solve the deviations for $A_{b}$ and $A_{FB}^{b}$ \cite{Chanowitz}.
Since the traditional 331 models are left-handed and the $Z-Z^{\prime }$
mixing is so weak ($\sim 10^{-3}$) they do not yield a contribution for
these asymmetries \cite{Martinez}. Another interesting possibility to
explain the discrepancy would be to modify the right-handed couplings of the
top quark, which enter in the correction of the $Zb\overline{b}$ vertex. It
could also generates deviations for $\left\vert V_{tb}\right\vert $, which
in turn may give us a hint about the mass generation mechanism for the
ordinary fermions. The 331 models with $N\neq 3$ might in principle be able
to explain such discrepancy and generate right-handed couplings for the
bottom and top quarks.

Furthermore, the introduction of mirror fermions permits in certain sense to
restore the chiral symmetry lost in the standard model, and in principle
could serve to solve the problem of strong CP violation \cite{Chang}. The
implementation of these models with more fermions for $N\neq 3$ requires a
more complex scalar sector that permits to generate CP violation in a
natural way.

There are some other features that neither SM nor their ordinary 331
extensions can explain at a cosmological level, such as the large scale
structure in the universe \cite{Volkas}, galactic halo \cite{Mohapatra}, and
gamma ray bursts \cite{Wong}. They suggest the existence of physics beyond
the ordinary 331 models. In many cases mirror fermions will be useful to
find solutions to these cosmological problems.

Finally, some additional motivations come from Grand Unified Theories
(GUT's). GUT's introduce some non-natural features such as the hierarchy
problem with the Higgs boson mass, because of the introduction of a new
scale (Grand Unification Scale) much higher than the weak scale, this is in
turn related with the \textquotedblleft grand desert\textquotedblright\ that
apparently exists between the GUT and electroweak scale. This fact motivates
the possibility of considering intermediate steps in the route from GUT to
EW scales. Some versions of the 331 models permits the chain of breaking GUT$%
\rightarrow 331\rightarrow SM$, while protecting the phenomenology from fast
proton decay \cite{Martinez2}.

The study of $\beta $ arbitrary is interesting because it permits a general
phenomenological analysis that could be reduced to the known cases when $%
\beta =-\sqrt{3},$ and $\beta =1/\sqrt{3}\ $\cite{Martinez Ochoa}, but can
also permit the study of other scenarios that could be the source for
solving some of the problems cited here.

Recently we have gotten constraints on 331 models by examining the scalar
sector \cite{331us}. In summary, these constraints are obtained by requiring
gauge invariance in the Yukawa sector and finding the possible vacuum
alignment structures that respect the symmetry breaking pattern and provides
the fermions and gauge bosons of the SM with the appropiate masses. By
applying gauge invariance to the Yukawa Lagrangian it is found that the
Higgs bosons should lie in either a triplet, antitriplet, singlet or sextet
representation of $SU\left( 3\right) _{L}$. On the other hand, cancellation
of chiral anomalies demands that the number of fermionic triplets and
antitriplets must be equal \cite{doff}. Moreover assuming the symmetry
breaking pattern 
\begin{eqnarray*}
SU\left( 3\right) _{c}\otimes SU\left( 3\right) _{L}\otimes U\left( 1\right)
_{X} &\rightarrow &SU\left( 3\right) _{c}\otimes SU\left( 2\right)
_{L}\otimes U\left( 1\right) _{Y}\rightarrow SU\left( 3\right) _{c}\otimes
U\left( 1\right) _{Q} \\
331 &\rightarrow &321\rightarrow 1
\end{eqnarray*}%
we see that one scalar triplet is necessary for the first symmetry breaking
and two scalar triplets for the second to give mass to the up and down
sectors of the SM. The possible vacuum alignments that obey this breaking
pattern as well as giving the appropiate masses in the second transition,
provide the value of the quantum number $X\ $in terms of $\beta $. Finally,
in some cases is necessary to introduce a scalar sextet to give masses to
all leptons.

In this paper we intend to make a general analysis of the fermionic spectrum
for $\beta $ arbitrary, by using the criteria of economy of the exotic
spectrum and the cancellation of anomalies. The scalar and vector sectors of
the model will be considered as well.

This paper is organized as follows. In Sec. \ref{sec:fermionic spectrum} we
describe the Fermion representations and find the restrictions over the
general fermionic structure based on the cancellation of anomalies. In Sec. %
\ref{sec:Higgspot} we show the scalar potential and the scalar spectrum for
three Higgs triplets with $\beta $ arbitrary. Sec. \ref{sec:vector spectrum}
developes the vector spectrum for $\beta $ arbitrary, and Sec. \ref%
{sec:Yang-Mills} shows the corresponding Yang-Mills Lagrangian. In Sec. \ref%
{sec:3 families} we write down the neutral and charged currents for the
three family version of the model with $\beta $ arbitrary. Sec. \ref{modelN4}
describes a new model with four families where one of them correspond to a
mirror family; from the vector-like structure of the model, we try to solve
the problem of the $b-$quark asymmetries, and generate ansatz for the
fermionic mass matrices. Finally, Sec. \ref{conclusions} is regarded for our
conclusions.

\section{Fermionic spectrum and anomalies with $\protect\beta $ arbitrary 
\label{sec:fermionic spectrum}}

\subsection{Fermion representations}

The fermion representations under $SU(3)_{c}\otimes $ $SU(3)_{L}\otimes
U(1)_{X}$ read

\begin{eqnarray}
\widehat{\psi }_{L} &=&\left\{ 
\begin{array}{c}
\widehat{q}_{L}:\left( \mathbf{3,3,}X_{q}^{L}\right) =\left( \mathbf{3,2,}%
X_{q}^{L}\right) \oplus \left( \mathbf{3,1,}X_{q}^{L}\right) , \\ 
\widehat{\ell }_{L}:\left( \mathbf{1,3,}X_{\ell }^{L}\right) =\left( \mathbf{%
1,2,}X_{\ell }^{L}\right) \oplus \left( \mathbf{1,1,}X_{\ell }^{L}\right) ,%
\end{array}
\right.  \notag \\
\widehat{\psi }_{L}^{\ast } &=&\left\{ 
\begin{array}{c}
\widehat{q}_{L}^{\ast }:\left( \mathbf{3,3}^{\ast }\mathbf{,-}%
X_{q}^{L}\right) =\left( \mathbf{3,2}^{\ast }\mathbf{,-}X_{q}^{L}\right)
\oplus \left( \mathbf{3,1,-}X_{q}^{L}\right) \mathbf{,} \\ 
\widehat{\ell }_{L}^{\ast }:\left( \mathbf{1,3}^{\ast }\mathbf{,-}X_{\ell
}^{L}\right) =\left( \mathbf{1,2}^{\ast }\mathbf{,-}X_{\ell }^{L}\right)
\oplus \left( \mathbf{1,1,-}X_{\ell }^{L}\right) ,%
\end{array}
\right.  \notag \\
\widehat{\psi }_{R} &=&\left\{ 
\begin{array}{c}
\widehat{q}_{R}:\left( \mathbf{3,1,}X_{q}^{R}\right) , \\ 
\widehat{\ell }_{R}:\left( \mathbf{1,1,}X_{\ell }^{R}\right) .%
\end{array}
\right.  \label{fermionrep}
\end{eqnarray}
The second equality comes from the branching rules $SU(2)_{L}\subset
SU(3)_{L}$. The $X_{p}$ refers to the quantum number associated with $%
U\left( 1\right) _{X}$. The generator of $U(1)_{X}$ conmute with the
matrices of $SU(3)_{L}$; hence, it should take the form $X_{p}\mathbf{I}%
_{3\times 3}$, the value of $X_{p}$ is related with the representations of $%
SU(3)_{L}$ and the anomalies cancellation. On the other hand, this fermionic
content shows that the left-handed multiplets lie in either the $\mathbf{3}$
or $\mathbf{3}^{\ast }$ representations.

\subsection{Chiral anomalies with $\protect\beta $ arbitrary}

The fermion spectrum in the SM consists of a set of three generations with
the same quantum numbers, the origin of these three generations is one of
the greatest puzzles of the model. On the other hand, the fermionic spectrum
of the 331 models must contain such generations, which can be fitted in
subdoublets $SU\left( 2\right) _{L}\subset SU\left( 3\right) _{L}$ according
to the structure given by Eq. (\ref{fermionrep}). Nevertheless, in such
models the number of fermion multiplets and their properties are related by
the condition of cancellation of anomalies. As a general starting point, we
could introduce sets of multiplets with different quantum numbers, it means
that each generation can be represented as a set of triplets with particles
of the SM plus exotic particles. Even in models of only one generation the
structure of the spectrum could be complex, appearing more than one triplet
with different quantum numbers \cite{fourteen}. These kind of models exhibit
a large quantity of free parameters and of exotic charges, such free
parameters increase rapidly when more than one generation is introduced, it
leads to a loss of predictibility in the sense that we have to resort to
phenomenological arguments to reduce the arbitrariness of the infinite
possible spectra. In the present work, we intend to study the 331 models
keeping certain generality but demanding a fermionic spectrum with a minimal
number of exotic particles. So we shall take all those models with $N$
leptonic generations and $M$ quark generations, by requiring to associate
only one lepton and one quark $SU\left( 3\right) _{L}\ $multiplet for each
generation, and at most one right-handed singlet associated with each
left-handed fermion. Based on these criteria we obtain the fermionic
spectrum (containing the SM spectrum) displayed in table \ref{tab:fercont}
for the quarks and leptons, where the definition of the electric charge Eq. (%
\ref{charge}), has been used demanding charges of $2/3$ and $-1/3$ to the up
and down-type quarks respectively, and charges of $-1,0$ for the charged and
neutral leptons, in order to ensure a realistic scenario. 
\begin{table}[tbp]
\begin{center}
\begin{equation*}
\begin{tabular}{||c||c||c||}
\hline\hline
$Quarks$ & $Q_{\psi }$ & $X_{\psi }$ \\ \hline\hline
\begin{tabular}{c}
$q_{L}^{(m)}=\left( 
\begin{array}{c}
U^{(m)} \\ 
D^{(m)} \\ 
J^{(m)}%
\end{array}
\right) _{L}:\mathbf{3}$ \\ 
\\ 
$U_{R}^{(m)}:\mathbf{1}$ \\ 
$D_{R}^{(m)}:\mathbf{1}$ \\ 
$J_{R}^{(m)}:\mathbf{1}$%
\end{tabular}
& 
\begin{tabular}{c}
$\left( 
\begin{array}{c}
\frac{2}{3} \\ 
-\frac{1}{3} \\ 
\frac{1}{6}-\frac{\sqrt{3}\beta }{2}%
\end{array}
\right) $ \\ 
\\ 
$\frac{2}{3}$ \\ 
$-\frac{1}{3}$ \\ 
$\frac{1}{6}-\frac{\sqrt{3}\beta }{2}$%
\end{tabular}
& 
\begin{tabular}{c}
\\ 
$X_{q^{(m)}}^{L}=\frac{1}{6}-\frac{\beta }{2\sqrt{3}}$ \\ 
\\ 
\\ 
$X_{U^{(m)}}^{R}=\frac{2}{3}$ \\ 
$X_{D^{(m)}}^{R}=-\frac{1}{3}$ \\ 
$X_{J^{(m)}}^{R}=\frac{1}{6}-\frac{\sqrt{3}\beta }{2}$%
\end{tabular}
\\ \hline\hline
\begin{tabular}{c}
$q_{L}^{(m^{\ast })}=\left( 
\begin{array}{c}
D^{(m^{\ast })} \\ 
-U^{(m^{\ast })} \\ 
J^{(m^{\ast })}%
\end{array}
\right) _{L}:\mathbf{3}^{\ast }$ \\ 
\\ 
$D_{R}^{(m^{\ast })}:\mathbf{1}$ \\ 
$U_{R}^{(m^{\ast })}:\mathbf{1}$ \\ 
$J_{R}^{(m^{\ast })}:\mathbf{1}$%
\end{tabular}
& 
\begin{tabular}{c}
$\left( 
\begin{array}{c}
-\frac{1}{3} \\ 
\frac{2}{3} \\ 
\frac{1}{6}+\frac{\sqrt{3}\beta }{2}%
\end{array}
\right) $ \\ 
\\ 
$-\frac{1}{3}$ \\ 
$\frac{2}{3}$ \\ 
$\frac{1}{6}+\frac{\sqrt{3}\beta }{2}$%
\end{tabular}
& 
\begin{tabular}{c}
\\ 
$X_{q^{(m^{\ast })}}^{L}=-\frac{1}{6}-\frac{\beta }{2\sqrt{3}}$ \\ 
\\ 
\\ 
$X_{D^{(m^{\ast })}}^{R}=-\frac{1}{3}$ \\ 
$X_{U^{(m^{\ast })}}^{R}=\frac{2}{3}$ \\ 
$X_{J^{(m^{\ast })}}^{R}=\frac{1}{6}+\frac{\sqrt{3}\beta }{2}$%
\end{tabular}
\\ \hline\hline
$Leptons$ & $Q_{\psi }$ & $X_{\psi }$ \\ \hline\hline
\begin{tabular}{c}
$\ell _{L}^{(n)}=\left( 
\begin{array}{c}
\nu ^{(n)} \\ 
e^{(n)} \\ 
E^{(n)}%
\end{array}
\right) _{L}:\mathbf{3}$ \\ 
\\ 
$\nu _{R}^{(n)}:\mathbf{1}$ \\ 
$e_{R}^{(n)}:\mathbf{1}$ \\ 
$E_{R}^{(n)}:\mathbf{1}$%
\end{tabular}
& 
\begin{tabular}{c}
$\left( 
\begin{array}{c}
0 \\ 
-1 \\ 
-\frac{1}{2}-\frac{\sqrt{3}\beta }{2}%
\end{array}
\right) $ \\ 
\\ 
$0$ \\ 
$-1$ \\ 
$-\frac{1}{2}-\frac{\sqrt{3}\beta }{2}$%
\end{tabular}
& 
\begin{tabular}{c}
\\ 
$X_{\ell ^{(n)}}^{L}=-\frac{1}{2}-\frac{\beta }{2\sqrt{3}}$ \\ 
\\ 
\\ 
$X_{\nu ^{(n)}}^{R}=0$ \\ 
$X_{e^{(n)}}^{R}=-1$ \\ 
$X_{E^{(n)}}^{R}=-\frac{1}{2}-\frac{\sqrt{3}\beta }{2}$%
\end{tabular}
\\ \hline\hline
\begin{tabular}{c}
$\ell _{L}^{(n^{\ast })}=\left( 
\begin{array}{c}
e^{(n^{\ast })} \\ 
-\nu ^{(n^{\ast })} \\ 
E^{(n^{\ast })}%
\end{array}
\right) _{L}:\mathbf{3}^{\ast }$ \\ 
\\ 
$e_{R}^{(n^{\ast })}:\mathbf{1}$ \\ 
$\nu _{R}^{(n^{\ast })}:\mathbf{1}$ \\ 
$E_{R}^{(n^{\ast })}:\mathbf{1}$%
\end{tabular}
& 
\begin{tabular}{c}
$\left( 
\begin{array}{c}
-1 \\ 
0 \\ 
-\frac{1}{2}+\frac{\sqrt{3}\beta }{2}%
\end{array}
\right) $ \\ 
\\ 
$-1$ \\ 
$0$ \\ 
$-\frac{1}{2}+\frac{\sqrt{3}\beta }{2}$%
\end{tabular}
& 
\begin{tabular}{c}
\\ 
$X_{\ell ^{(n^{\ast })}}^{L}=\frac{1}{2}-\frac{\beta }{2\sqrt{3}}$ \\ 
\\ 
\\ 
$X_{e^{(n^{\ast })}}^{R}=-1$ \\ 
$X_{\nu ^{(n^{\ast })}}^{R}=0$ \\ 
$X_{E^{(n^{\ast })}}^{R}=-\frac{1}{2}+\frac{\sqrt{3}\beta }{2}$%
\end{tabular}
\\ \hline\hline
\end{tabular}%
\end{equation*}%
\end{center}
\caption{\textit{Fermionic content of }$SU\left( 3\right) _{L}\otimes
U\left( 1\right) _{X}$\textit{\ obtained by requiring only one lepton and
one quark }$SU\left( 3\right) _{L}$\textit{\ multiplet for each generation,
and no more than one right-handed singlet for each right-handed field. The
structure of left-handed multiplets is the one shown in Eqs. (\protect\ref%
{fermionrep}, \protect\ref{notacion}). }$m$\textit{\ and }$n$\textit{\ label
the quark and lepton left-handed triplets respectively, while }$m^{\ast
},n^{\ast }$\textit{\ label the antitriplets, see Eq. (\protect\ref{notacion}%
).}}
\label{tab:fercont}
\end{table}
In general, it is possible to have in a single model any of the
representations described by Eq. (\ref{fermionrep}), where each multiplet
can transform differently. Indeed, in the most general case, each multiplet
can transform as 
\begin{equation}
\left\{ 
\begin{array}{l}
q_{L}^{\left( m\right) },q_{L}^{\left( m^{\ast }\right) }:\ m=\underset{3k\
triplets}{\underbrace{1,2,\ldots ,k}};\ m^{\ast }=\underset{3\left(
M-k\right) \ antitriplets}{\underbrace{k+1,k+2,\ldots ,M}} \\ 
\ell _{L}^{\left( n\right) },\ell _{L}^{\left( n^{\ast }\right) }\ \ \ \ :\
n=\underset{j\ triplets}{\underbrace{1,2,\ldots ,j}};\ n^{\ast }=\underset{%
N-j\ antitriplets}{\underbrace{j+1,j+2,\ldots ,N}}%
\end{array}%
\right.  \label{notacion}
\end{equation}%
where the first $3k$-th multiplets of quarks lie in the $\mathbf{3}$
representation while the latter $3\left( M-k\right) $ lie in the $\mathbf{3}%
^{\ast }$ representation for a total of $3M$ quark left-handed multiplets.
The factor $3$ in the number of quark left-handed multiplets owes to the
existence of three colors. Similarly the first $j$ left-handed multiplets of
leptons are taken in the representation $\mathbf{3}$ and the latter $\left(
N-j\right) $ are taken in the $\mathbf{3}^{\ast }$ representation, for a
total of $N$ leptonic left-handed multiplets.

Now we proceed to analize the restrictions over the fermionic structure of
Eq. (\ref{notacion}) from the criterion of cancellation of anomalies. When
we demand for the fermionic $SU(3)_{c}$ representations to be vector-like,
we are left with the following non-trivial triangular anomalies

\begin{eqnarray}
\left[ SU(3)_{c}\right] ^{2}\otimes U(1)_{X} &\rightarrow &A_{1}=\pm
3X_{q}^{L}-\sum_{\text{singlet}}X_{q}^{R}  \notag \\
\left[ SU(3)_{L}\right] ^{3} &\rightarrow &A_{2}=\frac{1}{2}A_{\alpha \beta
\gamma }  \notag \\
\left[ SU(3)_{L}\right] ^{2}\otimes U(1)_{X} &\rightarrow
&A_{3}=\sum\limits_{r}\left( \pm X_{\ell ^{(r)}}^{L}\right)
+3\sum\limits_{s}\left( \pm X_{q^{(s)}}^{L}\right) ,  \notag \\
\left[ Grav\right] ^{2}\otimes U(1)_{X} &\rightarrow
&A_{4}=3\sum\limits_{r}\left( \pm X_{\ell ^{(r)}}^{L}\right)
+9\sum\limits_{s}\left( \pm X_{q^{(s)}}^{L}\right)  \notag \\
&&-3\sum_{\text{singlet}}\left( X_{q}^{R}\right) -\sum_{\text{singlet}%
}\left( X_{\ell }^{R}\right) ,  \notag \\
\left[ U(1)_{X}\right] ^{3} &\rightarrow &A_{5}=3\sum\limits_{r}\left( \pm
X_{\ell ^{(r)}}^{L}\right) ^{3}+9\sum\limits_{s}\left( \pm
X_{q^{(s)}}^{L}\right) ^{3}  \notag \\
&&-3\sum_{\text{singlet}}\left( X_{q}^{R}\right) ^{3}-\sum_{\text{singlet}%
}\left( X_{\ell }^{R}\right) ^{3}  \label{anomalias}
\end{eqnarray}
where the sign $+$ or $-$ is chosen according to the representation $\mathbf{%
3}$ or $\mathbf{3}^{\ast }$. The condition of cancellation of these
anomalies imposes under some circumstances, relations between the values of $%
N,M,j,k$ and the $\beta $ parameter. Furthermore, the requirement for the
model to be $SU\left( 3\right) _{c}$ vector-like demands the presence of
right-handed quark singlets, while right-handed neutral lepton singlets are
optional.

\subsubsection{The $\left[ SU\left( 3\right) _{c}\right] ^{2}\otimes
U(1)_{X} $ anomaly}

When we take into account that the fermionic triplets in Eq. (\ref%
{fermionrep}) must contain the SM generations, i.e. they contain subdoublets 
$SU\left( 2\right) _{L}\subset SU\left( 3\right) _{L}$; we obtain relations
among the $X$ and $\beta $ numbers that cancel this anomaly. In table \ref%
{tab:fercont}, we write down these relations on the third column, by
assuming that the $SU\left( 2\right) _{L}$ subdoublets lies in the two upper
components of the triplets.

\subsubsection{The $\left[ SU\left( 3\right) _{L}\right] ^{3}$ anomaly}

The cancellation of the $\left[ SU(3)_{L}\right] ^{3}$ anomaly demands for
the number of $\widehat{\psi }_{L}$ multiplets to be the same as the number
of $\widehat{\psi }_{L}^{\ast }$ ones. Taking into account the number of
quark and lepton multiplets defined in Eq. (\ref{notacion}), we arrive to
the condition

\begin{equation*}
3k+j=3(M-k)+(N-j)
\end{equation*}
o rewriting it properly 
\begin{equation}
N-2j=-3(M-2k)\ ;\ 0\leq j\leq N\ ;\ 0\leq k\leq M.
\label{primera-restriccion}
\end{equation}
the first inequality expresses the fact that the models are limited from
representations in which all the left-handed multiplets of leptons transform
under $\mathbf{3}^{\ast }$ (when $j=0$),\ to representations in which all
left-handed lepton multiplets transform under $\mathbf{3}$ (when $j=N$). An
analogous situation appears for the quarks representations, that leads to
the second inequality.

\subsubsection{The $\left[ SU(3)_{L}\right] ^{2}\otimes U(1)_{X}$ anomaly}

Applying the definition in Eq. (\ref{notacion}), we make an explicit
separation between $\mathbf{3}$ and $\mathbf{3}^{\ast }$ representations,
from which this anomaly reads

\begin{equation*}
A_{3}=\sum\limits_{n=1}^{j}\left( X_{\ell ^{(n)}}^{L}\right)
+\sum\limits_{n^{\ast }=j+1}^{N}\left( -X_{\ell ^{(n^{\ast })}}^{L}\right)
+3\sum\limits_{m=1}^{k}\left( X_{q^{(m)}}^{L}\right) +3\sum\limits_{m^{\ast
}=k+1}^{M}\left( -X_{q^{(m^{\ast })}}^{L}\right) =0.
\end{equation*}
On the other hand, using the particle content of table \ref{tab:fercont},
the equation takes the form

\begin{equation}
-\frac{3}{2}M-\frac{3\sqrt{3}\beta }{2}(M-2k)=-\frac{3}{2}N+\frac{\sqrt{3}%
\beta }{2}(N-2j).  \label{segunda-restriccion}
\end{equation}

\subsubsection{The $\left[ Grav\right] ^{2}\otimes U(1)_{X}\ $anomaly}

Taking into account Eq. (\ref{notacion}), this anomaly takes the form

\begin{eqnarray*}
A_{4} &=&3\sum\limits_{n=1}^{j}\left( X_{\ell ^{(n)}}^{L}\right)
+3\sum\limits_{n^{\ast }=j+1}^{N}\left( -X_{\ell ^{(n^{\ast })}}^{L}\right)
+9\sum\limits_{m=1}^{k}\left( X_{q^{(m)}}^{L}\right) +9\sum\limits_{m^{\ast
}=k+1}^{M}\left( -X_{q^{(m^{\ast })}}^{L}\right) \\
&&-\sum\limits_{n=1}^{j}\left( Q_{\nu
^{(n)}}^{R}+Q_{e^{(n)}}^{R}+Q_{E^{(n)}}^{R}\right) -\sum\limits_{n^{\ast
}=j+1}^{N}\left( Q_{\nu ^{(n^{\ast })}}^{R}+Q_{e^{(n^{\ast
})}}^{R}+Q_{E^{(n^{\ast })}}^{R}\right) \\
&&-3\sum\limits_{m=1}^{k}\left( Q_{U^{(m)}}+Q_{D^{(m)}}+Q_{J^{(m)}}\right)
-3\sum\limits_{m^{\ast }=k+1}^{M}\left( Q_{U^{(m^{\ast })}}+Q_{D^{(m^{\ast
})}}+Q_{J^{(m^{\ast })}}\right) \\
&=&0\ ,
\end{eqnarray*}
where the leptonic right-handed charges can be present or absent. The
neutrino has null charge so that the presence (or absence) of right-handed
neutrinos does not affect the anomalies, but they are important when
choosing Yukawa terms for the masses. On the other hand, $e^{(n)}$ posseses
a charge $\left( -1\right) $, while $E^{(n)}$ and $E^{(n^{\ast })}\ $can in
general possess charges different from zero. We shall call them generically
charged leptons. Since charged singlets affect the anomalies, we should set
up a notation to specify whether we choose charged right-handed leptonic
singlets or not. Taking into account that we permit at most one right-handed
singlet per each left-handed fermion we define

\begin{equation}
\Theta _{\ell }\equiv \left\{ 
\begin{array}{c}
1\text{ \qquad\ \ for models\ with\ charged }\ell _{R} \\ 
0\text{ \qquad for models\ without\ charged }\ell _{R}%
\end{array}%
\right. ,  \label{paso}
\end{equation}%
it is applied to each right-handed leptonic charge, in such a way that the
cancellation of this anomaly leads to the condition 
\begin{eqnarray}
-\frac{3}{2}N+\frac{\sqrt{3}\beta }{2}(N-2j) &=&-j\Theta _{e^{(1)}}-j\left( 
\frac{1}{2}+\frac{\sqrt{3}\beta }{2}\right) \Theta _{E^{(1)}}  \notag \\
&&-(N-j)\Theta _{e^{(j+1)}}-(N-j)\left( \frac{1}{2}-\frac{\sqrt{3}\beta }{2}%
\right) \Theta _{E^{(j+1)}},  \label{tercera-restriccion}
\end{eqnarray}%
where we have replaced the values of $Q_{\psi },\ X_{\psi }$ given in table %
\ref{tab:fercont}. We should notice that the Eqs. (\ref{tercera-restriccion}%
) are relations about $\Theta _{\ell }$; therefore, they impose restrictions
over the possible choices of right-handed charged leptonic singlets.
Finally, from Eq. (\ref{tercera-restriccion}) and table \ref{tab:fercont},
we see that when the $E^{\left( n\right) }\ $or $E^{\left( n^{\ast }\right)
} $ fields are neutral (i.e. $\beta =\pm 1/\sqrt{3}$ for $E^{\left(
j+1\right) }$ and $E^{\left( 1\right) }$ respectively), the corresponding
singlets do not contribute to the equation of anomalies like in the case of
the neutrinos.

\subsubsection{$\left[ U(1)_{X}\right] ^{3}$ anomaly}

In this case we have: 
\begin{eqnarray*}
A_{5} &=&3\sum\limits_{n=1}^{j}\left( X_{\ell ^{(n)}}^{L}\right)
^{3}+3\sum\limits_{n^{\ast }=j+1}^{N}\left( -X_{\ell ^{(n^{\ast
})}}^{L}\right) ^{3}+9\sum\limits_{m=1}^{k}\left( X_{q^{(m)}}^{L}\right)
^{3}+9\sum\limits_{m^{\ast }=k+1}^{M}\left( -X_{q^{(m^{\ast })}}^{L}\right)
^{3} \\
&&-3\sum\limits_{m=1}^{k}\left[ \left( Q_{U^{(m)}}\right) ^{3}+\left(
Q_{D^{(m)}}\right) ^{3}+\left( Q_{J^{(m)}}\right) ^{3}\right]
-3\sum\limits_{m^{\ast }=k+1}^{M}\left[ \left( Q_{U^{(m^{\ast })}}\right)
^{3}+\left( Q_{D^{(m^{\ast })}}\right) ^{3}+\left( Q_{J^{(m^{\ast
})}}\right) ^{3}\right] \\
&&-\sum\limits_{n=1}^{j}\left[ \left( Q_{\nu ^{(n)}}^{R}\right) ^{3}+\left(
Q_{e^{(n)}}^{R}\right) ^{3}+\left( Q_{E^{(n)}}^{R}\right) ^{3}\right]
-\sum\limits_{n^{\ast }=j+1}^{N}\left[ \left( Q_{\nu ^{(n^{\ast
})}}^{R}\right) ^{3}+\left( Q_{e^{(n^{\ast })}}^{R}\right) ^{3}+\left(
Q_{E^{(n^{\ast })}}^{R}\right) ^{3}\right] \\
&=&0
\end{eqnarray*}
Using Eq. (\ref{paso}) and table \ref{tab:fercont}, we get 
\begin{eqnarray}
&&-\frac{3}{4}\left( \frac{1}{2}N+M\right) +\frac{3\sqrt{3}\beta }{8}(N-2j)-%
\frac{3\beta ^{2}}{4}\left( \frac{1}{2}N+M\right) +9\left( \frac{\beta }{%
\sqrt{3}}\right) ^{3}\left( \frac{N-2j}{24}-M+2k\right)  \label{NMbeta} \\
&=&-j\Theta _{e^{(1)}}-j\left( \frac{1}{2}+\frac{\sqrt{3}\beta }{2}\right)
^{3}\Theta _{E^{(1)}}-(N-j)\Theta _{e^{(j+1)}}-(N-j)\left( \frac{1}{2}-\frac{%
\sqrt{3}\beta }{2}\right) ^{3}\Theta _{E^{(j+1)}}  \notag
\end{eqnarray}
which arises as an additional condition for the presence of right-handed
charged leptonic singlets.

\subsection{General fermionic structure\label{sec:genfer}}

The Eqs. (\ref{primera-restriccion}), (\ref{segunda-restriccion}), (\ref%
{tercera-restriccion}) and (\ref{NMbeta}) appear as conditions that
guarantee the vanishing of all the anomalies, obtaining a set of four
equations (plus the two inequalities of Eq. (\ref{primera-restriccion}))
whose variables to solve for, are $N,M,j,k$ and $\beta $. Taking the two
equations (\ref{primera-restriccion}) and (\ref{segunda-restriccion}), we
find the following solutions 
\begin{equation}
N=M\ \ ;\ \ j+3k=2N.  \label{quinta-restriccion}
\end{equation}
this means that the number of left-handed quark multiplets ($3M$) must be
three times the number of left-handed leptonic multiplets ($N$). Moreover,
the number of leptonic triplets in the representation $\mathbf{3}$ ($j$)
plus the number of quark triplets in the representation $\mathbf{3}$ ($3k$)
must be twice the number of left-handed leptonic multiplets ($2N$) i.e. an
even number. In addition, we can find by combining the two of Eqs. (\ref%
{quinta-restriccion}), that the number of lepton and quarks left-handed
multiplets in the $\mathbf{3}^{\ast }$ representation must also be equal to $%
2N$. The solutions in Eqs. (\ref{quinta-restriccion}) are represented as
restrictions over the integer values of $j$ and $k$ according to the number
of left-handed multiplets ($4N$). Table \ref{tab:jksol}, illustrates some
particular cases. 

\begin{table}[tbp]
\begin{center}
\begin{equation*}
\begin{tabular}{||c||c||c||c||}
\hline\hline
$N$ & $0\leq j\leq N$ & $0\leq 3k\leq 3N$ & 
\begin{tabular}{c}
solution for \\ 
$j+3k=2N$%
\end{tabular}
\\ \hline\hline
1 & 0,1 & 0,3 & No solution. \\ \hline\hline
2 & 0,1,2 & 0,3,6 & $j=1;k=1$ \\ \hline\hline
3 & 0,1,2,3 & 0,3,6,9 & 
\begin{tabular}{c}
$j=0;k=2$ \\ 
$j=3;k=1$%
\end{tabular}
\\ \hline\hline
4 & 0,1,2,3,4 & 0,3,6,9,12 & $j=2;k=2$ \\ \hline\hline
5 & 0,1,2,3,4,5 & 0,3,6,9,12,15 & 
\begin{tabular}{c}
$j=1;k=3$ \\ 
$j=4;k=2$%
\end{tabular}
\\ \hline\hline
6 & 0,1,2,3,4,5,6 & 0,3,6,9,12,15,18 & 
\begin{tabular}{c}
$j=0;k=4$ \\ 
$j=3;k=3$ \\ 
$j=6;k=2$%
\end{tabular}
\\ \hline\hline
\end{tabular}%
\end{equation*}%
\end{center}
\caption{\textit{Solutions of Eqs. (\protect\ref{quinta-restriccion})
represented as restrictions on the number of lepton triplets }$\left(
j\right) $\textit{\ and}$\ $\textit{of quark triplets }$\left( 3k\right) $%
\textit{\ according to the number of left-handed multiplets }$\left(
4N\right) $\textit{.}}
\label{tab:jksol}
\end{table}

It is important to note that there are only some posible ways to choose the
number of triplets and antitriplets for a given number of multiplets.
Additionally, there is no solution for models with $N=1$ under the scheme of
using one multiplet per generation; so we have the extra condition $N\geq 2$%
. In this manner, the possible representations according to table \ref%
{tab:jksol}, depend on the number of multiplets $4N$, as it is shown in
table \ref{tab:jkrepres}. 
\begin{table}[tbp]
\begin{center}
\begin{tabular}{||c||c||}
\hline\hline
$N$ & Allowed representations \\ \hline\hline
2 & 
\begin{tabular}{|c|}
\hline
$\ell ^{(1)}:3$ \\ 
$\ell ^{(2)}:3^{\ast }$ \\ 
$q^{(1)}:3$ \\ 
$q^{(2)}:3^{\ast }$ \\ \hline
\end{tabular}
\\ \hline\hline
3 & 
\begin{tabular}{|l|}
\hline
$\ell ^{(1)},\ell ^{(2)},\ell ^{(3)}:3^{\ast }$ \\ 
$q^{(1)},q^{(2)}:3$ \\ 
$q^{(3)}:3^{\ast }$ \\ \hline
\end{tabular}
\begin{tabular}{|l|}
\hline
$\ell ^{(1)},\ell ^{(2)},\ell ^{(3)}:3$ \\ 
$q^{(3)}:3$ \\ 
$q^{(1)},q^{(2)}:3^{\ast }$ \\ \hline
\end{tabular}
\\ \hline\hline
4 & 
\begin{tabular}{|l|}
\hline
$\ell ^{(1)},\ell ^{(2)}:3$ \\ 
$\ell ^{(3)},\ell ^{(4)}:3^{\ast }$ \\ 
$q^{(1)},q^{(2)}:3$ \\ 
$q^{(3)},q^{(4)}:3^{\ast }$ \\ \hline
\end{tabular}
\\ \hline\hline
5 & 
\begin{tabular}{|l|}
\hline
$\ell ^{(5)}:3$ \\ 
$\ell ^{(1)},\ell ^{(2)},\ell ^{(3)},\ell ^{(4)}:3^{\ast }$ \\ 
$q^{(3)},q^{(4)},q^{(5)}:3$ \\ 
$q^{(1)},q^{(2)}:3^{\ast }$ \\ \hline
\end{tabular}
\begin{tabular}{|l|}
\hline
$\ell ^{(1)},\ell ^{(2)},\ell ^{(3)},\ell ^{(4)}:3$ \\ 
$\ell ^{(5)}:3^{\ast }$ \\ 
$q^{(1)},q^{(2)}:3$ \\ 
$q^{(3)},q^{(4)},q^{(5)}:3^{\ast }$ \\ \hline
\end{tabular}
\\ \hline\hline
6 & 
\begin{tabular}{c}
\begin{tabular}{|l|}
\hline
$%
\begin{tabular}{c}
$\ell ^{(1)},\ell ^{(2)},\ell ^{(3)},$ \\ 
$\ell ^{(4)},\ell ^{(5)},\ell ^{(6)}$%
\end{tabular}
\ :3^{\ast }$ \\ 
$q^{(1)},q^{(2)},q^{(5)},q^{(6)}:3$ \\ 
$q^{(3)},q^{(4)}:3^{\ast }$ \\ \hline
\end{tabular}
\begin{tabular}{|l|}
\hline
\begin{tabular}{c}
$\ell ^{(1)},\ell ^{(2)},\ell ^{(3)},$ \\ 
$\ell ^{(4)},\ell ^{(5)},\ell ^{(6)}$%
\end{tabular}
$:3$ \\ 
$q^{(3)},q^{(4)}:3$ \\ 
$q^{(1)},q^{(2)},q^{(5)},q^{(6)}:3^{\ast }$ \\ \hline
\end{tabular}
\\ 
\begin{tabular}{|l|}
\hline
$\ell ^{(1)},\ell ^{(2)},\ell ^{(3)}:3$ \\ 
$\ell ^{(4)},\ell ^{(5)},\ell ^{(6)}:3^{\ast }$ \\ 
$q^{(1)},q^{(2)},q^{(5)}:3$ \\ 
$q^{(3)},q^{(4)},q^{(6)}:3^{\ast }$ \\ \hline
\end{tabular}%
\end{tabular}
\\ \hline\hline
\end{tabular}%
\end{center}
\caption{\textit{Possible representations according to table \protect\ref%
{tab:jksol}. Each value of }$q^{\left( i\right) }$ represents three
left-handed quark multiplets because of the color factor.}
\label{tab:jkrepres}
\end{table}
We can see that models with $N=2$ are possible if the multiplets of quarks
and leptons transform in a different way. For $N=3$, we have two possible
solutions. In one of them all the lepton multiplets transform in the same
way, two of the quark multiplets transform the same and the other transform
as the conjugate. The second solution corresponds to the conjugate of the
first solution. For $N=4$ the quark and leptonic representations are
vector-like with respect to $SU\left( 3\right) _{L}$ as table \ref%
{tab:jkrepres} displays. In this case we will have one exotic fermion
family, $q^{(i)}$ and $l^{(i)}$; which might be a replication of the heavy
or light families of the SM. Such choice could be useful to generate new
ansatz about mass matrices for the fermions of the SM. In this way, it is
possible to add new exotic generations though not arbitrarily, but
respecting the conditions of the table \ref{tab:jkrepres}.

As for the solution (\ref{NMbeta}) with $N=M$, it can be rewritten as 
\begin{eqnarray}
\frac{3}{4}\left( 1+\beta ^{2}\right) \left[ -\frac{3}{2}N+\frac{\sqrt{3}%
\beta }{2}(N-2j)\right] &=&-j\Theta _{e^{(1)}}-j\left( \frac{1}{2}+\frac{%
\sqrt{3}\beta }{2}\right) ^{3}\Theta _{E^{(1)}}  \notag \\
&-&(N-j)\Theta _{e^{(j+1)}}-(N-j)\left( \frac{1}{2}-\frac{\sqrt{3}\beta }{2}%
\right) ^{3}\Theta _{E^{(j+1)}},
\end{eqnarray}%
and using (\ref{tercera-restriccion}), we find 
\begin{equation}
j\left( \Theta _{e^{(1)}}-\Theta _{E^{(1)}}\right) =(j-N)\left( \Theta
_{e^{(j+1)}}-\Theta _{E^{(j+1)}}\right)  \label{sexta-restriccion}
\end{equation}%
In this way, the solutions (\ref{tercera-restriccion}) and (\ref%
{sexta-restriccion}) represent restrictions over the singlet sector that are
related with the values of $N,j$ and $\beta $. All the possible combinations
of $\Theta _{\ell }$ that arise when the definition (\ref{paso}) is applied,
lead to the solutions summarized in table \ref{tab:tetasol}. 
\begin{table}[tbp]
\begin{center}
\scalebox{0.8}[0.8]{\ 
\begin{tabular}{||c||c||c||c||c||c||}
\hline\hline
\hspace{-0.2cm} 
\begin{tabular}{c}
$\mathbf{3}$ \\ 
$\Theta _{e^{(1)}}\;\Theta _{E^{(1)}}$\end{tabular}
\hspace{-0.2cm} & \hspace{-0.2cm} 
\begin{tabular}{c}
$\mathbf{3}^{\ast }$ \\ 
$\Theta _{e^{(j+1)}}\;\Theta _{E^{(j+1)}}$\end{tabular}
\hspace{-0.2cm} & 
\begin{tabular}{c}
Solution for \\ 
Eq. (\ref{tercera-restriccion})
\end{tabular}
& 
\begin{tabular}{c}
Solution for \\ 
Eq. (\ref{sexta-restriccion})
\end{tabular}
& 
\begin{tabular}{c}
Combined \\ 
solutions
\end{tabular}
& 
\begin{tabular}{c}
With conjugation \\ 
criterion
\end{tabular}
\\ \hline\hline
0\qquad 0 & 0\qquad 0 & $\beta =\left( \frac{N}{N-2j}\right) \sqrt{3}$ & $\forall $ $N$, $j$ & $\beta =\left( \frac{N}{N-2j}\right) \sqrt{3}$ & 
\begin{tabular}{c}
$\beta =\sqrt{3};$ $j=0$ \\ 
$\beta =-\sqrt{3};$ $j=N$\end{tabular}
\\ \hline\hline
0\qquad 0 & 0\qquad 1 & $\beta =\left( \frac{-2N-j}{j}\right) \frac{1}{\sqrt{3}}$ & $N=j$ & $\beta =-\sqrt{3}$ & \ding{55} \\ \hline\hline
0\qquad 0 & 1\qquad 0 & $\beta =\left( \frac{N+2j}{N-2j}\right) \frac{1}{\sqrt{3}}$ & $N=j$ & $\beta =-\sqrt{3}$ & \ding{55} \\ \hline\hline
0\qquad 0 & 1\qquad 1 & 
\begin{tabular}{c}
$\beta =-\sqrt{3};$ $\forall $ $j\neq 0$ \\ 
$\forall $ $\beta $; $j=0$\end{tabular}
& $\forall $ $N,$ $j$ & 
\begin{tabular}{c}
$\beta =-\sqrt{3};$ $\forall $ $j\neq 0$ \\ 
$\forall $ $\beta ;$ $j=0$\end{tabular}
& \ding{51} \\ \hline\hline
0\qquad 1 & 0\qquad 0 & $\beta =\left( \frac{3N-j}{N-j}\right) \frac{1}{\sqrt{3}}$ & $j=0;$ $\forall $ $N$ & $\beta =\sqrt{3}$ & \ding{55} \\ 
\hline\hline
0\qquad 1 & 0\qquad 1 & $N=0;$ $\forall $ $j,\beta $ & $N=0;$ $\forall $ $j\ 
$\ding{55} & $\forall $ $\beta $ & \ding{55} \\ \hline\hline
0\qquad 1 & 1\qquad 0 & $\beta =\left( \frac{N+j}{N-j}\right) \frac{1}{\sqrt{3}}$ & $N=2j$ & $\beta =\sqrt{3}$ & \ding{51} \\ \hline\hline
0\qquad 1 & 1\qquad 1 & $j=0;$ $\forall $ $N,\beta $ & $j=0;$ $\forall $ $N$
& $\forall $ $\beta $ & $\beta =-1/\sqrt{3}$ \\ \hline\hline
1\qquad 0 & 0\qquad 0 & $\beta =\left( \frac{3N-2j}{N-2j}\right) \frac{1}{\sqrt{3}}$ & $j=0;$ $\forall $ $N$ & $\beta =\sqrt{3}$ & \ding{55} \\ 
\hline\hline
1\qquad 0 & 0\qquad 1 & $\beta =\left( \frac{-2N+j}{j}\right) \frac{1}{\sqrt{3}}$ & $\;\;\quad \;N=2j\;\;\;\;$ & $\quad \beta =-\sqrt{3}\quad $ & \ding{51} \\ \hline\hline
1\qquad 0 & 1\qquad 0 & $\beta =\left( \frac{N}{N-2j}\right) \frac{1}{\sqrt{3}}$ & $N=0;$ $\forall $ $j\ $\ding{55} & $\beta =0$ & \ding{55} \\ 
\hline\hline
1\qquad 0 & 1\qquad 1 & 
\begin{tabular}{c}
$\beta =\frac{-1}{\sqrt{3}};$ $\forall $ $j\neq 0$ \\ 
$\forall $ $\beta $; $j=0$\end{tabular}
& $j=0;$ $\forall $ $N$ & $\forall $ $\beta $ & \ding{55} \\ \hline\hline
1\qquad 1 & 0\qquad 0 & 
\begin{tabular}{c}
$\beta =\sqrt{3};$ $\forall $ $j\neq N$ \\ 
$\forall $ $\beta $; $j=N$\end{tabular}
& $\forall $ $N$, $j$ & 
\begin{tabular}{c}
$\beta =\sqrt{3};$ $\forall $ $j\neq N$ \\ 
$\forall $ $\beta ;$ $j=N$\end{tabular}
& \ding{51} \\ \hline\hline
1\qquad 1 & 0\qquad 1 & $N=j;$ $\forall $ $\beta $ & $N=j$ & $\forall $ $\beta $ & $\beta =1/\sqrt{3}$ \\ \hline\hline
1\qquad 1 & 1\qquad 0 & 
\begin{tabular}{c}
$\beta =\frac{1}{\sqrt{3}};$ $\forall $ $j\neq N$ \\ 
$\forall $ $\beta $; $j=N$\end{tabular}
& $N=j$ & $\forall $ $\beta $ & \ding{55} \\ \hline\hline
1\qquad 1 & 1\qquad 1 & $\forall $ $N,j,\beta $ & $\forall $ $N,j$ & $\forall $ $\beta $ & 
\begin{tabular}{c}
if $j=0\Rightarrow \beta =-\sqrt{3}$ \\ 
if $j=N\Rightarrow \beta =\sqrt{3}$\end{tabular}
\\ \hline\hline
\end{tabular}
}
\end{center}
\caption{\textit{Solutions for Eqs. (\protect\ref{tercera-restriccion}) and (%
\protect\ref{sexta-restriccion}) that arise when all possible combinations
of }$\Theta _{\ell }$ defined by Eq. (\protect\ref{paso}) are taken. In the
last column, we mark with \ding{55} the cases that are ruled out by the
criterion of conjugation, while for the cases marked with \ding{51}, such
criterion does not give additional restrictions.}
\label{tab:tetasol}
\end{table}
Nevertheless, not all the 16 cases obtained correspond to physical
solutions. First of all $\beta =0$ is not permitted. Additionally, $N\geq 2$%
, from which the two solutions marked with \ding{55} on the fourth column of
table \ref{tab:tetasol}, are forbidden. On the other hand, there is another
important criterion to select possible physical models, which we shall call
the criterion of conjugation. The charged leptons are necessarily described
by Dirac's spinors, thus we should ensure for each charged lepton to include
its corresponding conjugate in the spectrum in order to build up the
corresponding Dirac Lagrangian. In the case of the exotic charged leptons
the conjugation criterion fixes their electric charges and so the possible
values of $\beta $, from which additional restrictions for the models are
obtained. As an example, for $\left( \Theta _{e^{(1)}},\ \Theta _{E^{(1)}},\
\Theta _{e^{(j+1)}},\ \Theta _{E^{(j+1)}}\right) =\left( 0,1,1,1\right) $
the cancellation of anomalies leads to $j=0\ $(see ninth row of table \ref%
{tab:tetasol})\ then, according to table \ref{tab:fercont}, the structure of
charged leptons is shown in table \ref{tab:0111}. Since the number of
leptons having non-zero charge must be even, one of the exotic leptons must
be neutral. Therefore, we have the following possibilities: \ding{202}
Demanding $E_{R}^{\left( n\right) }$ to be neutral we are led to $\beta =-1/%
\sqrt{3}$, now if we assume the scheme of conjugation $e_{L}^{(n^{\ast
})}\sim e_{R}^{(n^{\ast })};\ E_{L}^{(n^{\ast })}\sim E_{R}^{(n^{\ast })}$
no further restrictions are obtained. \ding{203} Assuming $E_{L}^{(n^{\ast
})}$ neutral, yields $\beta =1/\sqrt{3}$, but all possible combinations of
conjugation between the remaining charged fields are forbidden. For
instance, the scheme\ $E_{R}^{(n)}\sim e_{L}^{(n^{\ast })};\ E_{R}^{(n^{\ast
})}\sim e_{R}^{(n^{\ast })}$ yields\ $\beta =1/\sqrt{3}$ and $\beta =\sqrt{3}
$\ respectively, leading to a contradiction. \ding{204} Finally, for $%
E_{R}^{(n^{\ast })}$ neutral we find $\beta =1/\sqrt{3}$ and no consistent
conjugation structures are possible. In summary, for this singlet structure
the only value of $\beta $ consistent with the conjugation criterion is $%
\beta =1/\sqrt{3}$. This restriction should be added to the ones obtained
with cancellation of anomalies and yields the solutions shown in the ninth
row, last column of table \ref{tab:tetasol}. A similar procedure is done to
obtain the restrictions written in the last column of table \ref{tab:tetasol}%
. The cases marked with \ding{55} in the last column are forbidden, while
for the cases marked with \ding{51} the conjugation criterion provides no
further restrictions with respect to the ones obtained from cancellation of
anomalies.

\begin{table}[tbp]
\begin{center}
$%
\begin{tabular}{||c||c||}
\hline\hline
$Leptons$ & $Q_{\psi }$ \\ \hline\hline
\begin{tabular}{c}
$no\ triplets\ 3$ \\ 
$\nu _{R}^{(n)}$ \\ 
$E_{R}^{(n)}:1$%
\end{tabular}
& 
\begin{tabular}{c}
$no\ charge$ \\ 
$0$ \\ 
$-\frac{1}{2}-\frac{\sqrt{3}\beta }{2}$%
\end{tabular}
\\ \hline\hline
\begin{tabular}{c}
$\ell _{L}^{(n^{\ast })}=\left( 
\begin{array}{c}
e^{(n^{\ast })} \\ 
-\nu ^{(n^{\ast })} \\ 
E^{(n^{\ast })}%
\end{array}
\right) _{L}:3^{\ast }$ \\ 
\\ 
$e_{R}^{(n^{\ast })}:1$ \\ 
$\nu _{R}^{(n^{\ast })}:1$ \\ 
$E_{R}^{(n^{\ast })}:1$%
\end{tabular}
& 
\begin{tabular}{c}
$\left( 
\begin{array}{c}
-1 \\ 
0 \\ 
-\frac{1}{2}+\frac{\sqrt{3}\beta }{2}%
\end{array}
\right) $ \\ 
\\ 
$-1$ \\ 
$0$ \\ 
$-\frac{1}{2}+\frac{\sqrt{3}\beta }{2}$%
\end{tabular}
\\ \hline\hline
\end{tabular}
$%
\end{center}
\caption{Structure of leptons for the structure of singlets given by $\left(
\Theta _{e^{(1)}}\;,\Theta _{E^{(1)}},\ \Theta _{e^{(j+1)}}\;,\Theta
_{E^{(j+1)}}\right) =\left( 0,1,1,1\right) $.}
\label{tab:0111}
\end{table}

The solutions that survive in the table \ref{tab:tetasol} are combined with
the ones obtained in the tables \ref{tab:jksol}, \ref{tab:jkrepres} (or more
generally with Eqs. (\ref{quinta-restriccion})). The solutions that cancel
anomalies and fulfill the conjugation criterion are summarized in tables \ref%
{tab:tetarest1} and \ref{tab:tetarest2}. 

\begin{table}[tbp]
\begin{center}
\begin{tabular}{||c||c||c||c||c||}
\hline\hline
$\Theta _{e^{(1)}}$ & $\Theta _{E^{(1)}}$ & $\Theta _{e^{(j+1)}}$ & $\Theta
_{E^{(j+1)}}$ & Solution \\ \hline\hline
\ \ \thinspace 1\ \ \thinspace\  & \ \ \thinspace 0 \ \ \thinspace\  & \ \ \
\thinspace 0\ \ \thinspace\ \  & \ \ \ \thinspace \thinspace 1\ \ \
\thinspace \thinspace\  & $\quad \beta =-\sqrt{3}\quad $ \\ \hline\hline
0 & 1 & 1 & 0 & $\beta =\sqrt{3}$ \\ \hline\hline
\end{tabular}%
\end{center}
\caption{Solutions for \textit{$N=2j=2k\geq 2$}}
\label{tab:tetarest1}
\end{table}
\begin{table}[tbp]
\begin{center}
\begin{tabular}{||c||c||c||c||c||}
\hline\hline
$\Theta _{e^{(1)}}$ & $\Theta _{E^{(1)}}$ & $\Theta _{e^{(j+1)}}$ & $\Theta
_{E^{(j+1)}}$ & Solution \\ \hline\hline
0 & 0 & 0 & 0 & 
\begin{tabular}{c}
$\beta =\sqrt{3};$ $j=0$ \\ 
$\beta =-\sqrt{3};$ $j=N$%
\end{tabular}
\\ \hline\hline
0 & 0 & 1 & 1 & 
\begin{tabular}{c}
$\beta =-\sqrt{3};$ $\forall $ $j\neq 0$ \\ 
$\forall $ $\beta $; $j=0$%
\end{tabular}
\\ \hline\hline
1 & 1 & 0 & 0 & 
\begin{tabular}{c}
$\beta =\sqrt{3};$ $\forall $ $j\neq N$ \\ 
$\forall $ $\beta $; $j=N$%
\end{tabular}
\\ \hline\hline
0 & 1 & 1 & 1 & $j=0,\ \forall N,\ \beta =-1/\sqrt{3}$ \\ \hline\hline
1 & 1 & 0 & 1 & $j=N,\ \beta =1/\sqrt{3}$ \\ \hline\hline
1 & 1 & 1 & 1 & $%
\begin{array}{c}
\forall \beta ,\forall N,\ j\neq 0,N \\ 
\text{if\ }j=0\Rightarrow \beta =-\sqrt{3} \\ 
\text{if\ }j=N\Rightarrow \beta =\sqrt{3}%
\end{array}
$ \\ \hline\hline
\end{tabular}%
\end{center}
\caption{Solutions for \textit{$N=\frac{j+3k}2\geq 2$, $0\leq k\leq N$} }
\label{tab:tetarest2}
\end{table}

These solutions determine the fermionic structure of the model according to
the number of leptonic charged right-handed singlets. However, cancellation
of anomalies do not impose any restriction about the right-handed neutral
leptonic singlets. Table \ref{tab:tetarest1} only admits an even number of
left-handed leptonic multiplets ($N$), while table \ref{tab:tetarest2}
permits in principle any number of them as long as $N\geq 2$. It is observed
that there are models that fix the values of $\beta $, so that they are
possible only for certain values of the quantum numbers. However, in three
of the cases described in table \ref{tab:tetarest2}, there are solutions for 
$\beta $ arbitrary.

For the sake of completeness, we shall elaborate about the complications of
making an analysis of the most general case where we allow various
left-handed multiplets that transform in identical way with respect to $%
SU\left( 3\right) _{L}$, but with different quantum numbers with respect to $%
U\left( 1\right) _{X}$. In Eq. (\ref{notacion}), the number of left-handed
multiplets is enlarged to include the fact that each representation of $%
SU\left( 3\right) _{L}$ is formed by a subset of several left-handed
multiplets

\begin{equation}
\left\{ 
\begin{array}{l}
q^{(m)}:\text{ }m=\underbrace{1,....,m_{1}};\hspace{0.2cm}\underbrace{%
m_{1}+1,...,2m_{1}};.....;\underbrace{(k-1)m_{1}+1,....km_{1}}\text{ } \\ 
\hspace{1cm} 
\begin{tabular}{c}
$3m_{1}\text{ triplets.}$ \\ 
$1^{st}$ generation.%
\end{tabular}
\ \hspace{0.2cm} 
\begin{tabular}{c}
$3m_{1}\text{ triplets.}$ \\ 
$2^{nd}$ generation.%
\end{tabular}
\ ......... 
\begin{tabular}{c}
$3m_{1}\text{ triplets.}$ \\ 
$k-$th generation%
\end{tabular}
\\ 
\vspace{0.3cm} \\ 
q^{(m^{\ast })}:\text{ }m^{\ast }=\underbrace{km_{1}+1,....km_{1}+m_{1}^{%
\ast }};\hspace{0.2cm}\underbrace{km_{1}+m_{1}^{\ast
}+1,...,km_{1}+2m_{1}^{\ast }};... \\ 
\hspace{2cm} 
\begin{tabular}{c}
$3m_{1}^{\ast }\text{ antitriplets}$ \\ 
$(k+1)-$th generation%
\end{tabular}
\ \hspace{0.4cm} 
\begin{tabular}{c}
$3m_{1}^{\ast }\text{ antitriplets}$ \\ 
($k+2$)$-$th generation%
\end{tabular}
\ .... \\ 
\vspace{0.1cm} \\ 
\hspace{1.2cm}....;\underbrace{(M-1)m_{1}^{\ast }+1,....Mm_{1}^{\ast }} \\ 
\hspace{1.2cm}...... 
\begin{tabular}{c}
$3m_{1}^{\ast }\text{ antitriplets}$ \\ 
$M-$th generation%
\end{tabular}
\\ 
\vspace{0.3cm} \\ 
\ell ^{(n)}:\text{ }n=\underbrace{1,....n_{1}};\hspace{0.2cm}\underbrace{%
n_{1}+1,...,2n_{1}};.....;\underbrace{(j-1)n_{1}+1,....jn_{1}} \\ 
\hspace{0.8cm} 
\begin{tabular}{c}
$n_{1}\text{ triplets}$ \\ 
$1^{st}$ generation%
\end{tabular}
\ \hspace{0.2cm} 
\begin{tabular}{c}
$n_{1}\text{ triplets}$ \\ 
$2^{nd}$ generation%
\end{tabular}
\ ...... 
\begin{tabular}{c}
$n_{1}\text{ triplets}$ \\ 
$j-$th generation%
\end{tabular}
\\ 
\vspace{0.3cm} \\ 
\ell ^{(m^{\ast })}:\text{ }m^{\ast }=\underbrace{%
jn_{1}+1,....jn_{1}+n_{1}^{\ast }};\hspace{0.2cm}\underbrace{%
jn_{1}+n_{1}^{\ast }+1,...,jn_{1}+2n_{1}^{\ast }};... \\ 
\hspace{1.5cm} 
\begin{tabular}{c}
$n_{1}^{\ast }\text{ antitriplets}$ \\ 
$(j+1)-$th generation%
\end{tabular}
\ \hspace{0.4cm} 
\begin{tabular}{c}
$n_{1}^{\ast }\text{ antitriplets}$ \\ 
($j+2$)$-$ generation.%
\end{tabular}
\ ..... \\ 
\vspace{0.1cm} \\ 
\hspace{1.2cm}....;\underbrace{(N-1)n_{1}^{\ast }+1,....,Nn_{1}^{\ast }} \\ 
\hspace{1.2cm}..... 
\begin{tabular}{c}
$n_{1}^{\ast }\text{ antitriplets}$ \\ 
$N-$th generation%
\end{tabular}%
\end{array}
\right.  \label{notacion3}
\end{equation}

where $3m_{1},3m_{1}^{\ast },n_{1}$ and $n_{1}^{\ast }$ are the total number
of triplets and antitriplets for each generation of quarks (including the
color) and the total number of triplets and antitriplets for each generation
of leptons respectively. $k$ and $j$ are the number of generations of quarks
and leptons that transform according to $\mathbf{3}$, and $\left( M-k\right) 
$,$\ \left( N-j\right) $ are the number of generations under $\mathbf{3}%
^{\ast }$. In this way, the number of parameters is increased, having $%
N,M,n_{1},n_{1}^{\ast },m_{1},m_{1}^{\ast },j,k$ and $\beta $ as free
parameters, restricted by only four equations of cancellation of anomalies.
Since the number of triplets per generation (characterized by the indices $%
n_{1},n_{1}^{\ast },m_{1},m_{1}^{\ast }$) has no upper limit, it is always
possible to choose a convenient number of them to cancel anomalies, allowing
the entrance of an arbitrary number of exotic particles with no reasons but
purely phenomenological ones. Therefore, such models lose certain
naturalness which is precisely what we look for, when we build up a model
from basic principles with a minimum of free parameters.


\begin{table}[tbp]
\begin{center}
\begin{tabular}{||c||c||c||}
\hline\hline
$\Theta _{e^{(1)}}\;\Theta _{E^{(1)}}$ & $\Theta _{e^{(j+1)}}\;\Theta
_{E^{(j+1)}}$ & Solution \\ \hline\hline
0\qquad 0 & 0\qquad 0 & 
\begin{tabular}{c}
$\beta =\sqrt{3};$ $j=0,k=2$\ \ding{108} \\ 
$\beta =-\sqrt{3};$ $j=3,k=1$\ \ding{108}%
\end{tabular}
\\ \hline\hline
0\qquad 0 & 1\qquad 1 & 
\begin{tabular}{c}
$\beta =-\sqrt{3};$ $j=3,k=1$ \ding{117} \\ 
$\forall $ $\beta $; $j=0,k=2\ $\ding{111}%
\end{tabular}
\\ \hline\hline
0\qquad 1 & 1\qquad 1 & $\beta =-1/\sqrt{3};\ j=0,\ k=2$ \ding{79} \\ 
\hline\hline
1\qquad 1 & 0\qquad 0 & 
\begin{tabular}{c}
$\beta =\sqrt{3};$ $j=0,k=2$ \ding{71} \\ 
$\forall $ $\beta ;$ $j=3,k=1\ $\ding{111}%
\end{tabular}
\\ \hline\hline
1\qquad 1 & 0\qquad 1 & $\beta =1/\sqrt{3};\ j=3,\ k=1$\ \ding{79} \\ 
\hline\hline
1\qquad 1 & 1\qquad 1 & 
\begin{tabular}{c}
$\beta =-\sqrt{3};$ $j=0,k=2$ \ding{79} \\ 
$\beta =\sqrt{3};$ $j=3,k=1$ \ding{79}%
\end{tabular}
\\ \hline\hline
\end{tabular}%
\end{center}
\caption{\textit{Solutions for }$\protect\beta $ \textit{and the fermionic
structure with }$N=3$.}
\label{tab:N3sol}
\end{table}
For the case $N=3$ in table \ref{tab:tetarest2}, solutions exist only for $%
j=0$ or $3$ (see table \ref{tab:jksol}). These solutions are displayed in
table \ref{tab:N3sol}. It should be emphasized that the models without
leptonic right-handed singlets (marked with \ding{108}) are divided into two
according to the value of $j$ to be $0$ or $3$, which are precisely the
models discussed by Pleitez and Frampton \cite{ten, eleven}, where $\beta
=\pm \sqrt{3}$. The solutions marked with \ding{79} are not discarded by
anomalies nor conjugation, but lead to more than one right-handed singlet
for each left-handed field. On the other hand, the solutions marked with %
\ding{117} and \ding{71} gives no restriction on the number of right handed
leptonic singlets associated with $\mathbf{3}^{\ast }$ and $\mathbf{3}$
representations respectively. Finally, the solutions marked with \ding{111}
are the only ones that permit arbitrary values of $\beta $.

As for the two models with $\beta $ arbitrary, they exist only if leptonic
singlets associated with all the particles in either representation are
introduced. In the framework of these two solutions, the particular cases of
$\beta =\mp 1/\sqrt{3}$ are discussed by Long in Refs. \cite{twelve} and \cite{thirteen}
respectively.

It is interesting to notice that as well as the models of Pleitez, Frampton,
and Long, (with $\beta =\pm \sqrt{3},\pm 1/\sqrt{3}$) models with other
different values of $\beta $ arise. On the other hand, additional models
with $\beta =\pm \sqrt{3},\pm 1/\sqrt{3}$ but with different structures of
right-handed lepton singlets appear as well.

\section{Higgs Potential and spectrum for $\protect\beta $ arbitrary\label%
{sec:Higgspot}}

\subsection{Potential}

The scalar sector of the 331 models has also been studied in the literature 
\cite{331us, scalar331}. The most important features of the scalar potential
are \cite{331us}

\begin{itemize}
\item The scalars should lie in either the singlet, triplet, antitriplet, or
sextet representation of $SU\left( 3\right) _L$.

\item For the first transition $331\rightarrow 321$ we could have triplet,
antitriplet or sextet representations. The vacuum alignments for triplet and
antitriplet representations are indicated in table \ref{tab:tripalig} for $%
\beta \neq \pm 1/\sqrt{3}$. While for the sextet representation, the vacuum
alignment reads 
\begin{equation*}
\left\langle S^{ij}\right\rangle _0=\left[ 
\begin{array}{ccc}
0 & 0 & 0 \\ 
0 & 0 & 0 \\ 
0 & 0 & \nu _6%
\end{array}
\right]
\end{equation*}
These VEV's induce the masses of the exotic fermions.

\item In the second transition $321\rightarrow 31$, triplets, antitriplets
and sextets are also allowed. For the particular case of triplet (or
antitriplet) representations we get that pairs of solutions are obtained
according to the value of $\beta $. Both multiplets are necessary to give
masses to the quarks of type up and down respectively. So in the second
transition, we have to introduce two triplets (or antitriplets)$\;\rho $ and 
$\eta $ associated with each pair of solutions. We show in table \ref%
{tab:tripalig} the vacuum structure of this pair of triplets for $\beta \neq
\pm 1/\sqrt{3}$. On the other hand, the possible vacuum structures for the
second transition with Higgs sextets for $\beta \neq \pm 1/\sqrt{3},\pm 
\sqrt{3}$, are shown in table \ref{tab:sextalig}.

\item In some scenarios the Higgs sextet is necessary to give masses to all
leptons \cite{anomalias, doff}.
\end{itemize}


\begin{table}[tbh]
\begin{center}
$%
\begin{tabular}{|l|l|}
\hline
1$^{st}$ SSB$\; 
\begin{tabular}{|c|c|}
\hline
& $\beta \neq \pm \frac{1}{\sqrt{3}}$ \\ \hline
$\left\langle \chi \right\rangle _{0}$ & $\;\;\left( 
\begin{array}{c}
0 \\ 
0 \\ 
\nu _{\chi _{3}}%
\end{array}
\right) $ \\ \hline
$X_{\chi }$ & $\frac{\beta }{\sqrt{3}}$%
\end{tabular}
\ $ & 2$^{nd}$ SSB$\; 
\begin{tabular}{|c|c|}
\hline
$\left\langle \rho \right\rangle _{0}$ & $\left( 
\begin{array}{c}
0 \\ 
\nu _{\rho _{2}} \\ 
0%
\end{array}
\right) $ \\ \hline
$X_{\rho }$ & $\frac{1}{2}-\frac{\beta }{2\sqrt{3}}$ \\ \hline
$\left\langle \eta \right\rangle _{0}$ & $\left( 
\begin{array}{c}
\nu _{\eta _{1}} \\ 
0 \\ 
0%
\end{array}
\right) $ \\ \hline
$X_{\eta }$ & $-\frac{1}{2}-\frac{\beta }{2\sqrt{3}}$ \\ \hline
\end{tabular}
$ \\ \hline
\end{tabular}
\ $%
\end{center}
\caption{\textit{Vacuum alignments for the Higgs triplets neccesary to get
the SSB scheme: }$31\rightarrow 21\rightarrow 1$\textit{\ for }$\protect%
\beta \neq \pm 1/\protect\sqrt{3}$\textit{.\ In the case of Higgs
antitriplets, we find the same structure but replacing }$\protect\chi ,%
\protect\rho ,\protect\eta \rightarrow \protect\chi ^{\ast },\protect\rho %
^{\ast },\protect\eta ^{\ast }$\textit{.\ }}
\label{tab:tripalig}
\end{table}


\begin{table}[tbh]
\begin{center}
$%
\begin{tabular}{|c|c|}
\hline
& $\beta \neq \pm \frac{1}{\sqrt{3}};\pm \sqrt{3}$ \\ \hline
$\left\langle \rho ^{ij}\right\rangle _{0}$ & $\left( 
\begin{array}{ccc}
\nu _{1} & 0 & 0 \\ 
0 & 0 & 0 \\ 
0 & 0 & 0%
\end{array}
\right) \quad \left( 
\begin{array}{ccc}
0 & 0 & \nu _{3} \\ 
0 & 0 & 0 \\ 
\nu _{3} & 0 & 0%
\end{array}
\right) \quad \left( 
\begin{array}{ccc}
0 & 0 & 0 \\ 
0 & \nu _{4} & 0 \\ 
0 & 0 & 0%
\end{array}
\right) \quad \left( 
\begin{array}{ccc}
0 & 0 & 0 \\ 
0 & 0 & \nu _{5} \\ 
0 & \nu _{5} & 0%
\end{array}
\right) $ \\ \hline
$X_{\rho ^{ij}}$ & $-\frac{1}{2}-\frac{\beta }{2\sqrt{3}}\qquad \quad -\frac{%
1}{4}+\frac{\beta }{4\sqrt{3}}\quad \quad \qquad \frac{1}{2}-\frac{\beta }{2%
\sqrt{3}}\quad \quad \qquad \frac{1}{4}+\frac{\beta }{4\sqrt{3}}$ \\ \hline
\end{tabular}
$%
\end{center}
\caption{\textit{Vacuum alignments for the second SSB with Higgs sextets,
and for $\protect\beta \neq \pm 1/\protect\sqrt{3},\pm \protect\sqrt{3}$.}}
\label{tab:sextalig}
\end{table}
In the case of $\beta $ arbitrary (different from $\pm \sqrt{3},\ \pm 1/%
\sqrt{3}$), and taking a scalar content of three Higgs triplets, the most
general Higgs potential, renormalizable and $SU(3)_L\otimes U(1)_X$
invariant is \cite{331us}

\begin{eqnarray}
V_{higgs} &=&\mu _{1}^{2}\chi ^{i}\chi _{i}+\mu _{2}^{2}\rho ^{i}\rho
_{i}+\mu _{3}^{2}\eta ^{i}\eta _{i}+f\left( \chi _{i}\rho _{j}\eta
_{k}\varepsilon ^{ijk}+h.c.\right) +\lambda _{1}(\chi ^{i}\chi
_{i})^{2}+\lambda _{2}(\rho ^{i}\rho _{i})^{2}  \notag \\
&&+\lambda _{3}(\eta ^{i}\eta _{i})^{2}+\lambda _{4}\chi ^{i}\chi _{i}\rho
^{j}\rho _{j}+\lambda _{5}\chi ^{i}\chi _{i}\eta ^{j}\eta _{j}+\lambda
_{6}\rho ^{i}\rho _{i}\eta ^{j}\eta _{j}+\lambda _{7}\chi ^{i}\eta _{i}\eta
^{j}\chi _{j}  \notag \\
&&+\lambda _{8}\chi ^{i}\rho _{i}\rho ^{j}\chi _{j}+\lambda _{9}\eta
^{i}\rho _{i}\rho ^{j}\eta _{j}.  \label{gen b pot}
\end{eqnarray}
as it was mentioned above, in some models the choice of three triplets is
not enough to provide all leptons with masses \cite{anomalias, doff}. Hence,
an additional sextet is introduced. The choice of one of these solutions
depend on the fermionic sector to which we want to give masses. The
introduction of a sextet $S$, leads us to additional terms that should be
added to the Higgs potential of Eq. (\ref{gen b pot}) 
\begin{eqnarray}
V(S) &=&\mu _{5}^{2}S^{ij}S_{ij}+S^{ij}S_{ij}\left( \lambda _{15}\chi
^{k}\chi _{k}+\lambda _{16}\rho ^{k}\rho _{k}+\lambda _{17}\eta ^{k}\eta
_{k}\right) +\lambda _{18}\chi ^{i}S_{ij}S^{jk}\chi _{k}  \notag \\
&+&\lambda _{19}\rho ^{i}S_{ij}S^{jk}\rho _{k}+\lambda _{20}\eta
^{i}S_{ij}S^{jk}\eta _{k}+\lambda _{21}(S^{ij}S_{ij})^{2}+\lambda
_{22}S^{ij}S_{jk}S^{kl}S_{li}  \label{gen b pot2}
\end{eqnarray}

\subsection{Mass spectrum for $\protect\beta $ arbitrary \label{S masses}}

\begin{table}[tbp]
\begin{center}
$%
\begin{tabular}{|c|c|c|c|c|}
\hline
& $Q_{\Phi }$ & $Y_{\Phi }$ & $X_{\Phi }$ & $\left\langle \Phi \right\rangle
_{0}$ \\ \hline
$\chi =\left( 
\begin{array}{c}
\chi _{1}^{\pm Q_{1}} \\ 
\chi _{2}^{\pm Q_{2}} \\ 
\xi _{\chi }\pm i\zeta _{\chi }%
\end{array}
\right) $ & $\left( 
\begin{array}{c}
\pm \left( \frac{1}{2}+\frac{\sqrt{3}\beta }{2}\right) \\ 
\pm \left( -\frac{1}{2}+\frac{\sqrt{3}\beta }{2}\right) \\ 
0%
\end{array}
\right) $ & $\left( 
\begin{array}{c}
\pm \frac{\sqrt{3}\beta }{2} \\ 
\pm \frac{\sqrt{3}\beta }{2} \\ 
0%
\end{array}
\right) $ & $\frac{\beta }{\sqrt{3}}$ & $\left( 
\begin{array}{c}
0 \\ 
0 \\ 
\nu _{\chi }%
\end{array}
\right) $ \\ \hline
$\rho =\left( 
\begin{array}{c}
\rho _{1}^{\pm } \\ 
\xi _{\rho }\pm i\zeta _{\rho } \\ 
\rho _{3}^{\mp Q_{2}}%
\end{array}
\right) $ & $\left( 
\begin{array}{c}
\pm 1 \\ 
0 \\ 
\mp \left( -\frac{1}{2}+\frac{\sqrt{3}\beta }{2}\right)%
\end{array}
\right) $ & $\left( 
\begin{array}{c}
\pm \frac{1}{2} \\ 
\pm \frac{1}{2} \\ 
\,\mp \left( -\frac{1}{2}+\frac{\sqrt{3}\beta }{2}\right)%
\end{array}
\right) $ & $\frac{1}{2}-\frac{\beta }{2\sqrt{3}}$ & $\left( 
\begin{array}{c}
0 \\ 
\nu _{\rho } \\ 
0%
\end{array}
\right) $ \\ \hline
$\eta =\left( 
\begin{array}{c}
\xi _{\eta }\pm i\zeta _{\eta } \\ 
\eta _{2}^{\mp } \\ 
\eta _{3}^{\mp Q_{1}}%
\end{array}
\right) $ & $\left( 
\begin{array}{c}
0 \\ 
\mp 1 \\ 
\mp \left( \frac{1}{2}+\frac{\sqrt{3}\beta }{2}\right)%
\end{array}
\right) $ & $\left( 
\begin{array}{c}
\mp \frac{1}{2} \\ 
\mp \frac{1}{2} \\ 
\mp \left( \frac{1}{2}+\frac{\sqrt{3}\beta }{2}\right)%
\end{array}
\right) $ & $-\frac{1}{2}-\frac{\beta }{2\sqrt{3}}$ & $\left( 
\begin{array}{c}
\nu _{\eta } \\ 
0 \\ 
0%
\end{array}
\right) $ \\ \hline
\end{tabular}
\ $%
\end{center}
\caption{\textit{Quantum numbers of three scalar triplets for any $\protect%
\beta \neq \pm 1/\protect\sqrt{3}$.}}
\label{tab:quince}
\end{table}


In this section we analize the general case for $\beta $ arbitrary ($\beta
\neq \pm \sqrt{3},\pm 1/\sqrt{3}$). With three Higgs triplets, it is
obtained the potential given by Eq. (\ref{gen b pot}), which correspond to
the solution shown in table \ref{tab:tripalig} for $\beta \neq \pm 1/\sqrt{3}
$. In table \ref{tab:quince} we show the fields explicitly with their
corresponding charges, where $Q_{1}=\frac{1}{2}+\frac{\sqrt{3}\beta }{2}$
and $Q_{2}=-\frac{1}{2}+\frac{\sqrt{3}\beta }{2}$, refer to the electric
charge of the fields, which satisfy the property $Q_{1}-Q_{2}=1$. When we
apply the minimum conditions, the following relations are gotten 
\begin{eqnarray*}
\mu _{1}^{2} &=&-2\lambda _{1}\nu _{\chi }^{2}-\lambda _{4}\nu _{\rho
}^{2}-\lambda _{5}\nu _{\eta }^{2}-f\frac{\nu _{\eta }\nu _{\rho }}{\nu
_{\chi }}, \\
\mu _{2}^{2} &=&-2\lambda _{2}\nu _{\rho }^{2}-\lambda _{4}\nu _{\chi
}^{2}-\lambda _{6}\nu _{\eta }^{2}-f\frac{\nu _{\eta }\nu _{\chi }}{\nu
_{\rho }}, \\
\mu _{3}^{2} &=&-2\lambda _{3}\nu _{\eta }^{2}-\lambda _{5}\nu _{\chi
}^{2}-\lambda _{6}\nu _{\rho }^{2}-f\frac{\nu _{\rho }\nu _{\chi }}{\nu
_{\eta }}.
\end{eqnarray*}

\noindent and we replace them again in the scalar potential to find the
physical spectrum of the fields and their masses. From the second
derivatives with respect to the fields, we obtain the mass matrices $%
M_{\zeta \zeta }^{2}$ for the imaginary sector, $M_{\xi \xi }^{2}$ for the
scalar real sector and three decoupled matrices $M_{\phi }^{2}$ for the
scalar charged sector.

In order to obtain the eigenvalues and eigenvectors we shall suppose that
there is a strong hierarchy between the scales of the first and the second
transition, from which it is natural to assume 
\begin{equation}
\left\langle \chi \right\rangle _{0}\gg \left\langle \rho \right\rangle
_{0},\left\langle \eta \right\rangle _{0}\Rightarrow \left\vert \nu _{\chi
}\right\vert >>\left\vert \nu _{\rho }\right\vert ,\left\vert \nu _{\eta
}\right\vert  \label{hierarchy1}
\end{equation}%
In addition, since some of the Higgs bosons of the first transition are
proportional to $f\nu _{\chi }$ we shall make the assumption 
\begin{equation}
\left\vert f\right\vert \approx \left\vert \nu _{\chi }\right\vert
\label{hierarchy2}
\end{equation}%
where $f$ is the trilinear coupling constant defined in the scalar potential
Eq. (\ref{gen b pot}). This assumption prevents the introduction of another
scale different from the ones defined by the two transitions. In our
approach we shall keep only the matrix elements that are quadratic in $\nu
_{\chi }$ i.e. the terms proportional to $\nu _{\chi }^{2},\ f\nu _{\chi }$
unless otherwise is indicated. Under these approximations, the mass matrices
and eigenvalues are written in explicit form, in Eqs.(\ref{C1})-(\ref%
{charged Higgses}) in appendix \ref{ap:beta}. Summarizing we get all the
scalar bosons described in table \ref{tab:dieciseis}.$\ $

\begin{table}[tbp]
\begin{center}
$%
\begin{tabular}{||c||c||c||}
\hline\hline
Charged scalars & Square masses & Feature \\ \hline\hline
$\phi _2^0\simeq -\zeta _\chi $ & $M_{\phi _2^0}^2=0$ & 
\begin{tabular}{l}
Goldstone \\ 
associated with $Z_\mu ^{\prime }$%
\end{tabular}
\\ \hline\hline
$\phi _3^0\simeq S_\beta \zeta _\rho -C_\beta \zeta _\eta $ & $M_{\phi
_3^0}^2=0$ & 
\begin{tabular}{l}
Goldstone \\ 
associated with $Z_\mu $%
\end{tabular}
\\ \hline\hline
$\phi _1^{\pm }=S_\beta \rho _1^{\pm }-C_\beta \eta _2^{\pm },$ & $M_{\phi
_1^{\pm }}^2=0$ & 
\begin{tabular}{l}
Goldstone \\ 
associated with $W_\mu ^{\pm }$%
\end{tabular}
\\ \hline\hline
$\phi _2^{\pm Q_1}\simeq -\chi _1^{\pm Q_1}$ & $M_{\phi _2^{\pm }}^2=0$ & $%
\begin{tabular}{l}
Goldstone \\ 
associated with $K_\mu ^{\pm Q_1}$%
\end{tabular}
\ $ \\ \hline\hline
$\phi _3^{\pm Q_2}\simeq -\chi _2^{\pm Q_2}$ & $M_{\phi _3^{\pm }}^2=0$ & $%
\begin{tabular}{l}
Goldstone \\ 
associated with $K_\mu ^{\pm Q_2}$%
\end{tabular}
$ \\ \hline\hline
$h_1^0\simeq C_\beta \zeta _\rho +S_\beta \zeta _\eta $ & $M_{h_1^0}^2\simeq
-2f\nu _\chi \left( \frac{\nu _\eta }{\nu _\rho }+\frac{\nu _\rho }{\nu
_\eta }\right) $ & Higgs \\ \hline\hline
$h_3^0\simeq S_\beta \xi _\rho +C_\beta \xi _\eta $ & $M_{h_3^0}^2\simeq
\frac 8{\nu _\eta ^2+\nu _\rho ^2}\left[ \lambda _2\nu _\rho ^4+2\lambda
_6\nu _\rho ^2\nu _\eta ^2+\lambda _3\nu _\eta ^4\right] $ & Higgs \\ 
\hline\hline
$h_4^0\simeq -C_\beta \xi _\rho +S_\beta \xi _\eta $ & $M_{h_4^0}^2\simeq
-2f\nu _\chi \left( \frac{\nu _\eta }{\nu _\rho }+\frac{\nu _\rho }{\nu
_\eta }\right) $ & Higgs \\ \hline\hline
$h_5^0\simeq \xi _\chi $ & $M_{h_5^0}^2\simeq 8\lambda _1\nu _\chi ^2$ & 
Higgs \\ \hline\hline
$h_1^{\pm Q_1}=\eta _3^{\pm Q_1}$ & $M_{h_1^{\pm }}^2\simeq \lambda _7\nu
_\chi ^2-f\nu _\chi \frac{\nu _\rho }{\nu _\eta }$ & Higgs \\ \hline\hline
$h_2^{\pm }=C_\beta \rho _1^{\pm }+S_\beta \eta _2^{\pm }$ & $M_{h_2^{\pm
}}^2\simeq -f\nu _\chi \left( \frac{\nu _\eta }{\nu _\rho }+\frac{\nu _\rho 
}{\nu _\eta }\right) $ & Higgs \\ \hline\hline
$h_3^{\pm Q_2}=\rho _3^{\pm Q_2}$ & $M_{h_3^{\pm }}^2\simeq \lambda _8\nu
_\chi ^2-f\nu _\chi \frac{\nu _\eta }{\nu _\rho }$ & Higgs \\ \hline\hline
\end{tabular}
$%
\end{center}
\caption{\textit{Spectrum of scalars for $\protect\beta \neq \pm 1/\protect%
\sqrt{3},\mathit{\pm }\protect\sqrt{3}.$}}
\label{tab:dieciseis}
\end{table}

\section{Vector spectrum with $\protect\beta $ arbitrary\label{sec:vector
spectrum}}

The gauge bosons associated with the $SU(3)_{L}$ group transform according
to the adjoint representation and are written in the form 
\begin{equation}
W_{\mu }=W_{\mu }^{\alpha }G_{\alpha }=\frac{1}{2}\left[ 
\begin{array}{ccc}
W_{\mu }^{3}+\frac{1}{\sqrt{3}}W_{\mu }^{8} & \sqrt{2}W_{\mu }^{+} & \sqrt{2}%
K_{\mu }^{Q_{1}} \\ 
\sqrt{2}W_{\mu }^{-} & -W_{\mu }^{3}+\frac{1}{\sqrt{3}}W_{\mu }^{8} & \sqrt{2%
}K_{\mu }^{Q_{2}} \\ 
\sqrt{2}K_{\mu }^{-Q_{1}} & \sqrt{2}K_{\mu }^{-Q_{2}} & -\frac{2}{\sqrt{3}}%
W_{\mu }^{8}%
\end{array}%
\right] .  \label{3}
\end{equation}%
Therefore, the electric charge takes the general form 
\begin{equation}
Q_{W}\rightarrow \left[ 
\begin{array}{ccc}
0 & 1 & \frac{1}{2}+\frac{\sqrt{3}\beta }{2} \\ 
-1 & 0 & -\frac{1}{2}+\frac{\sqrt{3}\beta }{2} \\ 
-\frac{1}{2}-\frac{\sqrt{3}\beta }{2} & \frac{1}{2}-\frac{\sqrt{3}\beta }{2}
& 0%
\end{array}%
\right] .  \label{4}
\end{equation}%
As for the gauge field associated with $U(1)_{X},$ it is represented as$\ 
\mathbf{B}_{\mu }=B_{\mu }\mathbf{I}_{3\times 3}$ which is a singlet under $%
SU(3)_{L},$ and has no electric charge. From the previous expressions we see
that three gauge fields with charges equal to zero are obtained, and in the
basis of mass eigenstates they correspond to the photon, $Z$ and $Z^{\prime
} $. Moreover, there are two fields with charges $\pm 1$ associated with $%
W^{\pm }$, as well as four fields with charges that depend on the choice of $%
\beta $ (denoted by $K^{\pm Q_{1}}$ and $K^{\pm Q_{2}}$). Demanding that the
model contains no exotic charges in this sector, is equivalent to setting up 
$\beta =-1/\sqrt{3}$ \cite{twelve}, and $\beta =1/\sqrt{3}\ $\cite{331us}.
It is important to take into account the scalar sector and the symmetry
breakings to fix this quantum number, which in turn determine the would-be
Goldstone bosons associated with the gauge fields, with the same electric
charge of the gauge fields that are acquiring mass in the different scales
of breakdown.

\subsection{Charged sector}

The masses for $W^{\pm },K^{\pm Q_{1}},K^{\pm Q_{2}}$ charged gauge fields
read 
\begin{equation*}
M^2_{W^{\pm }}=\frac{g^{2}}{2}\left( \nu _{\rho }^{2}+\nu _{\eta
}^{2}\right) , M^2_{K^{\pm Q_{1}}}=\frac{g^{2}}{2}\left( \nu _{\chi
}^{2}+\nu _{\eta }^{2}\right) , M^2_{K^{\pm Q_{2}}}=\frac{g^{2}}{2}\left(
\nu _{\chi }^{2}+\nu _{\rho }^{2}\right)
\end{equation*}
where the terms proportional to $\nu _{\chi }^{2}$ acquire heavy masses of
the order of the first symmetry breaking. The other fields acquire a mass
proportional to the electroweak scale and correspond to the gauge fields $%
W^{\pm }$. The mass eigenstates are given by 
\begin{equation}
W_{\mu }^{\pm }=\frac{1}{\sqrt{2}}\left( W_{\mu }^{1}\mp iW_{\mu
}^{2}\right) \ ;\ K_{\mu }^{\pm Q_{1}}=\frac{1}{\sqrt{2}}\left( W_{\mu
}^{4}\mp iW_{\mu }^{5}\right) \ ;\ K_{\mu }^{\pm Q_{2}}=\frac{1}{\sqrt{2}}%
\left( W_{\mu }^{6}\mp iW_{\mu }^{7}\right)
\end{equation}

\subsection{Neutral sector}

The mass matrix is given in appendix \ref{massneutralgauge}. This matrix has
null determinant corresponding to the mass of the photon. After the proper
rotation the mass eigenstates become 
\begin{eqnarray}
A_{\mu } &=&S_{W}W_{\mu }^{3}+C_{W}\left( \beta T_{W}W_{\mu }^{8}+\sqrt{%
1-\beta ^{2}T_{W}^{2}}B_{\mu }\right) ,  \notag \\
Z_{\mu }^{\prime } &=&-\sqrt{1-\beta ^{2}\left( T_{W}\right) ^{2}}W_{\mu
}^{8}+\beta T_{W}B_{\mu },  \notag \\
Z_{\mu } &=&C_{W}W_{\mu }^{3}-S_{W}\left( \beta T_{W}W_{\mu }^{8}+\sqrt{%
1-\beta ^{2}T_{W}^{2}}B_{\mu }\right) ,
\end{eqnarray}%
The corresponding eigenvalues are 
\begin{equation}
M_{A_{\mu }}^{2}=0\ ;\ M_{Z_{\mu }^{\prime }}^{2}\simeq \frac{2\left[
g^{2}+\beta ^{2}g^{\prime 2}\right] }{3}\nu _{\chi }^{2}\ ;\ M_{Z_{\mu
}}^{2}\simeq \frac{g^{2}}{2}\left[ \frac{g^{2}+\left( 1+\beta ^{2}\right)
g^{\prime 2}}{g^{2}+\beta ^{2}g^{\prime 2}}\right] \left( \nu _{\rho
}^{2}+\nu _{\eta }^{2}\right)
\end{equation}%
where the Weinberg angle is defined (in terms of $\beta $) as: 
\begin{equation}
S_{W}\equiv \sin \theta _{W}=\frac{g^{\prime }}{\sqrt{g^{2}+\left( 1+\beta
^{2}\right) g^{\prime 2}}}\ .
\end{equation}%
and $g,$ $g^{\prime }$ correspond to the coupling constants of the groups $%
SU(3)_{L}$ and $U(1)_{X}$, respectively. Further, a small mixing between the 
$Z_{\mu }$ and $Z_{\mu }^{\prime }$ could occur getting 
\begin{eqnarray}
Z_{1\mu } &=&Z_{\mu }\cos \theta +Z_{\mu }^{\prime }\sin \theta \ \ ;\ \
Z_{2\mu }=-Z_{\mu }\sin \theta +Z_{\mu }^{\prime }\cos \theta ,  \notag \\
\tan \theta &=&\frac{1}{\Lambda +\sqrt{\Lambda ^{2}+1}}\ \ ;\ \ \Lambda =%
\frac{-2S_{W}C_{W}^{2}g^{\prime 2}\nu _{\chi }^{2}+\frac{3}{2}%
S_{W}T_{W}^{2}g^{2}\left( \nu _{\eta }^{2}+\nu _{\rho }^{2}\right) }{%
gg^{\prime }T_{W}^{2}\left[ 3S_{W}^{2}\beta \left( \nu _{\eta }^{2}+\nu
_{\rho }^{2}\right) +C_{W}^{2}\left( \nu _{\eta }^{2}-\nu _{\rho
}^{2}\right) \right] }
\end{eqnarray}

It is interesting to notice that from the definition of the charge in Eq. (%
\ref{charge}), we obtain a matching condition among the coupling constants,
that in turn leads to the following expression 
\begin{equation*}
\frac{g^{\prime 2}}{g^2}=\frac{S_W^2}{1-S_W^2\left( 1+\beta ^2\right) }
\end{equation*}
By running the Weinberg angle through renormalization group equations, we
can find a scale to which a singularity of this quotient appears. In some
models and for certain values of $\beta $, this pole could appear at the TeV
scale \cite{Martinez Pleitez}.

We point out that when $\beta =-\sqrt{3}$, we get the same definitions and
diagonalizations of the model of Pleitez and Frampton \cite{ten, eleven}.
The $\beta $ parameter can be written explicitly in terms of the exotic
charges as $\beta =\left( 2Q_{1}-1\right) /\sqrt{3}=\left( 2Q_{2}+1\right) /%
\sqrt{3}$. From which it is obtained that in general$\ Q_{1}-Q_{2}=1$, so
independently of the model the difference in charges between the charged
gauge fields will be equal to the unity.

\section{Yang-Mills Couplings\label{sec:Yang-Mills}}

In general, the Yang-Mills Lagrangian for $SU(3)_{L}\times U(1)_{Y}\ $is\
given\ by 
\begin{equation}
\mathfrak{L}_{YM}=-\frac{1}{4}W_{\mu \nu }^{i}W_{i}^{\mu \nu }+\frac{1}{2}%
gf_{ijk}W_{\mu \nu }^{i}W^{\mu j}W^{\nu k}-\frac{g^{2}}{4}%
f^{ijk}f_{ilm}W_{\mu }^{j}W^{\mu k}W_{\nu }^{l}W^{\nu m}-\frac{1}{4}B_{\mu
\nu }B^{\mu \nu },
\end{equation}
where $W_{i}^{\mu \nu }=\partial ^{\mu }W_{i}^{\nu }-\partial ^{\nu
}W_{i}^{\mu }$. After writing this Lagrangian in terms of the mass
eigenstates, the cubic couplings read 
\begin{eqnarray*}
\mathfrak{L}_{cubic} &=&e\left\{ \left[ r-p\right] ^{\mu }g^{\alpha \nu }+%
\left[ p-q\right] ^{\nu }g^{\alpha \mu }+\left[ q-r\right] ^{\alpha }g^{\nu
\mu }\right\} A_{\nu }W_{\alpha }^{+}W_{\mu }^{-} \\
&&+Q_{1}e\left\{ \left[ r-p\right] ^{\mu }g^{\alpha \nu }+\left[ p-q\right]
^{\nu }g^{\alpha \mu }+\left[ q-r\right] ^{\alpha }g^{\nu \mu }\right\}
A_{\nu }K_{\alpha }^{+Q_{1}}K_{\mu }^{-Q_{1}} \\
&&+Q_{2}e\left\{ \left[ r-p\right] ^{\mu }g^{\alpha \nu }+\left[ p-q\right]
^{\nu }g^{\alpha \mu }+\left[ q-r\right] ^{\alpha }g^{\nu \mu }\right\}
A_{\nu }K_{\alpha }^{+Q_{2}}K_{\mu }^{-Q_{2}} \\
&&+gC_{W}\left\{ \left[ p-q\right] ^{\mu }g^{\alpha \nu }+\left[ q-r\right]
^{\alpha }g^{\nu \mu }+\left[ r-p\right] ^{\nu }g^{\alpha \mu }\right\}
Z_{\mu }W_{\alpha }^{+}W_{\nu }^{-} \\
&&+\left[ \frac{gC_{W}}{2}+\frac{\left( 2Q_{1}-1\right) eT_{W}}{2}\right]
\left\{ \left[ p-q\right] ^{\mu }g^{\alpha \nu }+\left[ q-r\right] ^{\alpha
}g^{\nu \mu }+\left[ r-p\right] ^{\nu }g^{\alpha \mu }\right\} Z_{\mu
}K_{\alpha }^{+Q_{1}}K_{\nu }^{-Q_{1}} \\
&&+\left[ \frac{-gC_{W}}{2}+\frac{\left( 2Q_{2}+1\right) eT_{W}}{2}\right]
\left\{ \left[ p-q\right] ^{\mu }g^{\alpha \nu }+\left[ q-r\right] ^{\alpha
}g^{\nu \mu }+\left[ r-p\right] ^{\nu }g^{\alpha \mu }\right\} Z_{\mu
}K_{\alpha }^{+Q_{2}}K_{\nu }^{-Q_{2}} \\
&&+\frac{\sqrt{3}g}{2}\sqrt{1-\beta ^{2}T_{W}^{2}}\left\{ \left[ q-p\right]
^{\mu }g^{\alpha \nu }+\left[ r-q\right] ^{\alpha }g^{\nu \mu }+\left[ p-r%
\right] ^{\nu }g^{\alpha \mu }\right\} Z_{\mu }^{\prime }K_{\alpha
}^{+Q_{1}}K_{\nu }^{-Q_{1}} \\
&&+\frac{\sqrt{3}g}{2}\sqrt{1-\beta ^{2}T_{W}^{2}}\left\{ \left[ q-p\right]
^{\mu }g^{\alpha \nu }+\left[ r-q\right] ^{\alpha }g^{\nu \mu }+\left[ p-r%
\right] ^{\nu }g^{\alpha \mu }\right\} Z_{\mu }^{\prime }K_{\alpha
}^{+Q_{2}}K_{\nu }^{-Q_{2}}.
\end{eqnarray*}
In passing to the space of momenta we associate $\partial _{\mu }=-ip_{\mu }$
and the following assignments of momenta: $p_{\mu }$ for the positively
charged fields $W_{\nu }^{+}$, $K_{\nu }^{+Q_{1}}$, and $K_{\nu }^{+Q_{2}}$; 
$q_{\mu }$ for the negatively charged fields i.e. for $W_{\nu }^{-}$, $%
K_{\nu }^{-Q_{1}}$, $K_{\nu }^{-Q_{2}}$; finally, $r_{\mu }$ for the neutral
fields $A_{\nu }$, $Z_{\nu }$, $Z_{\nu }^{\prime }$. It is assumed that all
the momenta enter to the vertex of interaction and that the sum of them
vanishes. Note that the coupling $Z_{\mu }^{\prime }W^{\pm }W^{\mp }$ does
not appear at tree level, because of the form of the $f_{ijk}$ structure
constant.

Further, the quartic hermitian couplings are

\begin{center}
\begin{eqnarray*}
\mathfrak{L}_{quartic} &=&g^{2}W_{\alpha }^{-}W_{\beta }^{+}\left\{ -%
\mathfrak{g}_{1}^{\alpha \delta \gamma \beta }\left[ W_{\gamma
}^{+}W_{\delta }^{-}+K_{\gamma }^{-Q_{2}}K_{\delta }^{+Q_{2}}\right] +%
\mathfrak{g}_{2}^{\alpha \beta \gamma \delta }\left[ S_{W}^{2}A_{\gamma
}A_{\delta }+C_{W}^{2}Z_{\gamma }Z_{\delta }+S_{W}C_{W}A_{\gamma }Z_{\delta }%
\right] \right. \\
&&\left. -\frac{1}{2}\mathfrak{g}_{3}^{\beta \delta \alpha \gamma }K_{\gamma
}^{-Q_{1}}K_{\delta }^{+Q_{1}}\right\} \\
&&+g^{2}K_{\alpha }^{-Q_{1}}K_{\beta }^{+Q_{1}}\left\{ -\mathfrak{g}%
_{1}^{\alpha \delta \gamma \beta }K_{\gamma }^{+Q_{1}}K_{\delta }^{-Q_{1}}+%
\mathfrak{g}_{2}^{\alpha \beta \gamma \delta }\left[ S_{W}^{2}Q_{1}^{2}A_{%
\gamma }A_{\delta }+\frac{C_{W}^{2}}{4}\left( \sqrt{3}\beta
T_{W}^{2}-1\right) ^{2}Z_{\gamma }Z_{\delta }\right. \right. \\
&&-\frac{S_{W}C_{W}Q_{1}}{2}\left( \sqrt{3}\beta T_{W}^{2}-1\right)
A_{\gamma }Z_{\delta }-\frac{\sqrt{3}S_{W}Q_{1}}{2}\sqrt{1-\beta
^{2}T_{W}^{2}}A_{\gamma }Z_{\delta }^{\prime }+\frac{3}{4}\left( 1-\beta
^{2}T_{W}^{2}\right) Z_{\gamma }^{\prime }Z_{\delta }^{\prime } \\
&&\left. \left. +\frac{\sqrt{3}C_{W}}{4}\sqrt{1-\beta ^{2}T_{W}^{2}}\left( 
\sqrt{3}\beta T_{W}^{2}-1\right) Z_{\gamma }Z_{\delta }^{\prime }\right]
\right\} \\
&&+g^{2}K_{\alpha }^{-Q_{2}}K_{\beta }^{+Q_{2}}\left\{ -\mathfrak{g}%
_{1}^{\alpha \delta \gamma \beta }K_{\gamma }^{+Q_{2}}K_{\delta }^{-Q_{2}}+%
\mathfrak{g}_{2}^{\alpha \beta \gamma \delta }\left[ S_{W}^{2}Q_{2}^{2}A_{%
\gamma }A_{\delta }+\frac{C_{W}^{2}}{4}\left( \sqrt{3}\beta
T_{W}^{2}+1\right) ^{2}Z_{\gamma }Z_{\delta }\right. \right. \\
&&-\frac{S_{W}C_{W}Q_{2}}{2}\left( \sqrt{3}\beta T_{W}^{2}+1\right)
A_{\gamma }Z_{\delta }-\frac{\sqrt{3}S_{W}Q_{2}}{2}\sqrt{1-\beta
^{2}T_{W}^{2}}A_{\gamma }Z_{\delta }^{\prime } \\
&&\left. \left. +\frac{\sqrt{3}C_{W}}{4}\sqrt{1-\beta ^{2}T_{W}^{2}}\left( 
\sqrt{3}\beta T_{W}^{2}+1\right) Z_{\gamma }Z_{\delta }^{\prime }+\frac{3}{4}%
\left( 1-\beta ^{2}T_{W}^{2}\right) Z_{\gamma }^{\prime }Z_{\delta }^{\prime
}\right] -\frac{1}{2}\mathfrak{g}_{3}^{\beta \delta \alpha \gamma }K_{\gamma
}^{-Q_{2}}K_{\delta }^{+Q_{2}}\right\}
\end{eqnarray*}

\begin{eqnarray}
&&-\frac{\sqrt{3}g^2}{2\sqrt{2}}\sqrt{1-\beta ^2T_W^2}\mathfrak{g}_2^{\alpha
\beta \gamma \delta }K_\alpha ^{+Q_2}K_\beta ^{-Q_1}W_\gamma ^{+}Z_\delta
^{\prime }+h.c.  \notag \\
+\hspace{-0.3cm} &&\frac{g^2S_W}{2\sqrt{2}}\left[ \left( Q_1+Q_2\right) 
\mathfrak{g}_2^{\alpha \beta \gamma \delta }+\mathfrak{g}_3^{\beta \delta
\alpha \gamma }-\mathfrak{g}_1^{\alpha \delta \gamma \beta }\right] K_\alpha
^{+Q_2}K_\beta ^{-Q_1}W_\gamma ^{+}A_\delta +h.c.  \notag \\
&&\hspace{-0.3cm}-\frac{g^2C_W}{\sqrt{2}}\left[ \frac{\left( \sqrt{3}\beta
T_W^2+1\right) }2\mathfrak{g}_2^{\alpha \beta \gamma \delta }+\mathfrak{g}%
_1^{\alpha \delta \gamma \beta }\right] K_\alpha ^{+Q_2}K_\beta
^{-Q_1}W_\gamma ^{+}Z_\delta +h.c,
\end{eqnarray}
\end{center}

where $\mathfrak{g}_{1}^{\alpha \delta \gamma \beta }=-2g^{\alpha \delta
}g^{\gamma \beta }+g^{\alpha \beta }g^{\gamma \delta }+g^{\alpha \gamma
}g^{\beta \delta },$ $\mathfrak{g}_{2}^{\alpha \beta \gamma \delta
}=-2g^{\alpha \beta }g^{\gamma \delta }+g^{\alpha \gamma }g^{\beta \delta
}+g^{\alpha \delta }g^{\beta \gamma }$ and $\mathfrak{g}_{3}^{\beta \delta
\alpha \gamma }=-2g^{\beta \delta }g^{\alpha \gamma }+g^{\alpha \beta
}g^{\gamma \delta }+g^{\alpha \delta }g^{\beta \gamma }.$ These results are
in agreement with Ref. \cite{Tavares} for $\beta =\sqrt{3}$ and Ref. \cite%
{Toscano} for $\beta =\frac{-1}{\sqrt{3}}$.

\section{Model for $N=3$ with $\protect\beta $ arbitrary\label{sec:3
families}}

In section (\ref{sec:genfer}), we found that if $\beta \neq \pm \sqrt{3},\pm
1/\sqrt{3}$, there are only two solutions for the fermionic structure when $%
N=3$ (the ones marked with \ding{111} in table \ref{tab:N3sol}), where the
solutions are the complex conjugate of each other. Then, we take the option
with $j=3$ (three lepton triplets), $k=1$ (one quark triplet and two
antitriplets) valid for all $\beta $. Applying this solution to the
fermionic content given in table \ref{tab:fercont}, we obtain the fermionic
spectrum given in table \ref{tab:espectro}. We should notice that for this
solution, we must introduce right-handed leptonic singlets associated with
each left-handed lepton ($\Theta _{e^{(1)}}=\Theta _{E^{(1)}}=1$) for Eqs. (%
\ref{tercera-restriccion}) and (\ref{sexta-restriccion}) to be accomplished
ensuring the vanishing of the anomalies.


\begin{table}[tbp]
\begin{center}
\begin{equation*}
\begin{tabular}{||c||c||c||}
\hline\hline
$representation$ & $Q_\psi $ & $X_\psi $ \\ \hline\hline
$\ 
\begin{tabular}{c}
$q_{m^{*}L}=\left( 
\begin{array}{c}
d,s \\ 
-u,-c \\ 
J_1,J_2%
\end{array}
\right) _L\mathbf{3}^{*}$ \\ 
\\ 
\\ 
$d_{m^{*}R}=d_R,s_R:\mathbf{1}$ \\ 
$u_{m^{*}R}=u_R,c_R:\mathbf{1}$ \\ 
$J_{m^{*}R}=J_{1R},J_{2R}:\mathbf{1}$%
\end{tabular}
$ & 
\begin{tabular}{c}
$\left( 
\begin{array}{c}
-\frac 13 \\ 
\frac 23 \\ 
\frac 16+\frac{\sqrt{3}\beta }2%
\end{array}
\right) $ \\ 
\\ 
$-\frac 13$ \\ 
$\frac 23$ \\ 
$\frac 16+\frac{\sqrt{3}}2\beta $%
\end{tabular}
& 
\begin{tabular}{c}
\\ 
$X_{q^{(m)}}^L=-\frac 16-\frac \beta {2\sqrt{3}}$ \\ 
\\ 
\\ 
$X_{u^{(m)}}^R=-\frac 13$ \\ 
$X_{d^{(m)}}^R=\frac 23$ \\ 
$X_{J^{(m)}}^R=\frac 16+\frac{\sqrt{3}}2\beta $%
\end{tabular}
\\ \hline\hline
\begin{tabular}{c}
$q_{3L}=\left( 
\begin{array}{c}
t \\ 
b \\ 
J_3%
\end{array}
\right) _L:\mathbf{3}$ \\ 
\\ 
$u_{3R}=b_R:\mathbf{1}$ \\ 
$d_{3R}=t_R:\mathbf{1}$ \\ 
$J_{3R}=J_{3R}:\mathbf{1}$%
\end{tabular}
& 
\begin{tabular}{c}
$\left( 
\begin{array}{c}
\frac 23 \\ 
-\frac 13 \\ 
\frac 16-\frac{\sqrt{3}\beta }2%
\end{array}
\right) $ \\ 
\\ 
$-\frac 13$ \\ 
$\frac 23$ \\ 
$\frac 16-\frac{\sqrt{3}\beta }2$%
\end{tabular}
& 
\begin{tabular}{c}
\\ 
$X_{q^{(3)}}^L=\frac 16-\frac \beta {2\sqrt{3}}$ \\ 
\\ 
\\ 
$X_b^R=-\frac 13$ \\ 
$X_t^R=\frac 23$ \\ 
$X_{J_3}^R=\frac 16-\frac{\sqrt{3}\beta }2$%
\end{tabular}
\\ \hline\hline
\begin{tabular}{c}
$\ell _{jL}=\left( 
\begin{array}{c}
\nu _e,\nu _\mu ,\nu _\tau \\ 
e^{-},\mu ^{-},\tau ^{-} \\ 
E_1^{-Q_1},E_2^{-Q_1},E_3^{-Q_1}%
\end{array}
\right) _L:\mathbf{3}$ \\ 
\\ 
$\left( e_j^{-}\right) _R=e^{-},\mu ^{-},\tau _R^{-}:\mathbf{1}$ \\ 
$E_j^{-Q_1}=E_1^{-Q_1},E_2^{-Q_1},E_3^{-Q_1}:\mathbf{1}$%
\end{tabular}
& 
\begin{tabular}{c}
$\left( 
\begin{array}{c}
0 \\ 
-1 \\ 
-\frac 12-\frac{\sqrt{3}\beta }2%
\end{array}
\right) $ \\ 
\\ 
$-1$ \\ 
$-\frac 12-\frac{\sqrt{3}\beta }2$%
\end{tabular}
& 
\begin{tabular}{c}
\\ 
$X_{\ell ^{(m)}}^L=-\frac 12-\frac \beta {2\sqrt{3}}$ \\ 
\\ 
\\ 
$X_{e^{(m)}}^R=-1$ \\ 
$X_{E_m}^R=-\frac 12-\frac{\sqrt{3}\beta }2$%
\end{tabular}
\\ \hline\hline
\end{tabular}%
\end{equation*}%
\end{center}
\caption{\textit{Fermionic content for $N=3$ with\ }$\protect\beta \ $%
\textit{arbitrary. $m^{\ast }=\mathit{1,2\ }$and$\ \mathit{j=1,2,3}$\textit{.%
}}}
\label{tab:espectro}
\end{table}

\subsection{Neutral and charged currents}

The Dirac Lagrangian contains the couplings between gauge bosons and
fermions, given by

\begin{equation*}
\mathcal{L}_{\mathcal{F}}=i\overline{\psi _{L}}\partial \hspace{-0.2cm}/\psi
_{L}\mp g\overline{\psi _{L}}W\hspace{-0.4cm}/\hspace{0.2cm}\psi _{L}\mp
g^{\prime }\overline{\psi _{L}}B\hspace{-0.2cm}/X_{p}^{L}\psi _{L}+i%
\overline{\psi _{R}}\partial \hspace{-0.2cm}/\psi _{R}-g^{\prime }\overline{%
\psi _{R}}B\hspace{-0.2cm}/X_{p}^{R}\psi _{R}
\end{equation*}%
where the sign is chosen according to the representation $\mathbf{3}$ or\ $%
\mathbf{3}^{\ast }$ respectively. Since the mass matrices mix the quarks
among each other, the mass basis is different from the gauge basis. So when
we write the Lagrangian in terms of mass eigenstates we get

\begin{eqnarray}
\mathcal{L}_{q} &=&eQ_{q_{j}}\overline{\mathbf{Q}}_{j}\gamma _{\mu }A^{\mu }%
\mathbf{Q}_{j}+\frac{g}{C_{W}}\overline{\mathbf{Q}_{j}}\gamma _{\mu }\left[
T_{3}P_{L}-Q_{q_{j}}S_{W}^{2}\right] Z^{\mu }\mathbf{Q}_{j}  \notag \\
&&+\frac{g}{\sqrt{2}}\overline{d_{iL}}\gamma _{\mu }(U_{\theta j}^{i})^{\ast
}W^{\mu -}u_{jL}+\frac{g}{\sqrt{2}}\overline{u_{jL}}\gamma _{\mu }W^{\mu
+}U_{\theta j}^{i}d_{iL}  \notag \\
&&+\frac{g^{\prime }}{2T_{W}}\overline{q_{m^{\ast }}}\gamma _{\mu }\left[
\left( 2T_{8}+\beta Q_{q_{m^{\ast }}}T_{W}^{2}\Lambda _{1}\right)
P_{L}+2\beta Q_{q_{m^{\ast }}}T_{W}^{2}P_{R}\right] Z^{\mu \prime
}q_{m^{\ast }}  \notag \\
&&+\frac{g^{\prime }}{2T_{W}}\overline{q_{3}}\gamma _{\mu }\left[ \left(
-2T_{8}+\beta Q_{q_{3}}T_{W}^{2}\Lambda _{2}\right) P_{L}+2\beta
Q_{q_{3}}T_{W}^{2}P_{R}\right] Z^{\mu \prime }q_{3}  \notag \\
&&-\frac{g}{\sqrt{2}}\overline{d_{iL}}\gamma _{\mu }(U_{\theta j}^{i})^{\ast
}\delta _{j}^{n^{\ast }}K^{\mu ^{-Q_{1}}}U_{\phi n^{\ast }}^{m^{\ast
}}J_{m^{\ast }L}-\frac{g}{\sqrt{2}}\overline{J_{m^{\ast }L}}\gamma _{\mu
}(U_{\phi n^{\ast }}^{m^{\ast }})^{\ast }K^{\mu ^{+Q_{1}}}\delta _{n^{\ast
}}^{j}U_{\theta j}^{i}d_{iL}  \notag \\
&&+\frac{g}{\sqrt{2}}\overline{J_{3L}}\gamma _{\mu }K^{\mu ^{-Q_{1}}}t_{L}+%
\frac{g}{\sqrt{2}}\overline{t_{L}}\gamma _{\mu }K^{\mu ^{+Q_{1}}}J_{3L} 
\notag \\
&&+\frac{g}{\sqrt{2}}\overline{u_{n^{\ast }L}}\gamma _{\mu }K^{\mu
^{-Q_{2}}}U_{\phi n^{\ast }}^{m^{\ast }}J_{m^{\ast }L}+\frac{g}{\sqrt{2}}%
\overline{J_{m^{\ast }L}}\gamma _{\mu }\left( U_{\phi n^{\ast }}^{m^{\ast
}}\right) ^{\ast }K^{\mu ^{+Q_{2}}}u_{n^{\ast }L}  \notag \\
&&+\frac{g}{\sqrt{2}}\overline{J_{3L}}\gamma _{\mu }K^{\mu ^{-Q_{2}}}\delta
_{3}^{j}U_{\theta j}^{i}d_{iL}+\frac{g}{\sqrt{2}}\overline{d_{iL}}\gamma
_{\mu }\left( U_{\theta j}^{i}\right) ^{\ast }\delta _{j}^{3}K^{\mu
^{+Q_{2}}}J_{3L}.  \label{L-quarks}
\end{eqnarray}%
The couplings associated with $A^{\mu }$ and $Z^{\mu }$ have been written in
a SM-like notation i.e. $\mathbf{Q}_{j}$ with $j=1,2,3$ refers to triplets
in the $\mathbf{3}$ representation associated with the three generations of
quarks.

On the other hand, the couplings of the exotic gauge bosons with the two
former families, are different from the ones involving the third famliy. It
is because the third family transforms differently (see table \ref%
{tab:espectro}). Consequently, there are terms where only the components $%
m^{*},n^{*}=1,2$ are summed, leaving the third one in a term apart. $%
q_{m^{*}}$ refers to the two triplets of quarks with $q_{1,2}$ in the $%
\mathbf{3}^{*}$ representation and $q_3$ in the $\mathbf{3}$ representation. 
$Q_{q_{m^{*}}}$ are their electric charges shown in table \ref{tab:espectro}%
.\ We define $\Lambda _1\equiv diag\left( -1,1/2,2\right) $ and $\Lambda
_2\equiv diag(\frac 12,-1,2)$. We also have used the projectors $%
P_{R,L}=(1\pm \gamma _5)/2$. Flavor mixings appear owing to the charged
gauge bosons $W^\mu ,K^{\mu ^{\pm Q_1}}$ and $K^{\mu ^{\pm Q_2}}$, where the
CKM matrix $U_\theta \ $has been defined with the usual mixing angles $%
\theta _i$ of the SM and a matrix $U_\phi $ with a mixing angle $\phi _c$
associated with the exotic quarks $J_1$ and $J_2$ (the quark $J_3$ is
decoupled in the mass matrices because of its different electric charge).
Eq. (\ref{L-quarks}) includes the SM couplings properly.

As for the leptons, we have for the three families

\begin{eqnarray}
\mathcal{L}_{\ell _{m}} &=&eQ_{\ell _{j}}\overline{\ell _{j}}\gamma _{\mu
}A^{\mu }\ell _{j}+\frac{g}{C_{W}}\overline{\ell _{j}}\gamma _{\mu }\left[
T_{3}P_{L}-Q_{\ell _{j}}S_{W}^{2}\right] Z^{\mu }\ell _{j}  \notag \\
&&+\frac{g}{\sqrt{2}}\overline{\nu _{jL}}\gamma _{\mu }W^{\mu +}e_{jL}^{-}+%
\frac{g}{\sqrt{2}}\overline{e_{jL}^{-}}\gamma _{\mu }W^{\mu -}\nu _{jL} 
\notag \\
&&+\frac{g^{\prime }}{2T_{W}}\overline{\ell _{j}}\gamma _{\mu }\left[ \left(
-2T_{8}-\beta T_{W}^{2}\Lambda _{3}\right) P_{L}+2Q_{\ell _{j}}\beta
T_{W}^{2}P_{R}\right] Z^{\mu \prime }\ell _{j}  \notag \\
&&+\frac{g}{\sqrt{2}}\overline{\nu _{jL}}\gamma _{\mu }K^{\mu
^{+Q_{1}}}E_{jL}^{-Q_{1}}+\frac{g}{\sqrt{2}}\overline{E_{jL}^{-Q_{1}}}\gamma
_{\mu }K^{\mu ^{-Q_{1}}}\nu _{jL}  \notag \\
&&+\frac{g}{\sqrt{2}}\overline{e_{jL}^{-}}\gamma _{\mu }K^{\mu
^{+Q_{2}}}E_{jL}^{-Q_{1}}+\frac{g}{\sqrt{2}}\overline{E_{jL}^{-Q_{1}}}\gamma
_{\mu }K^{\mu ^{-Q_{2}}}e_{jL}^{-},  \label{L-leptones}
\end{eqnarray}%
with $\ell _{j\text{ }}$denoting the leptonic triplets shown in table \ref%
{tab:espectro}, and with $Q_{\ell _{j}}$ denoting their electric charges,
finally $\Lambda _{3}=diag(1,1,2Q_{1}).$

\section{Model with $N=4$ and $\protect\beta =-\frac{1}{\protect\sqrt{3}}$%
\label{modelN4}}

\begin{table}[!tbh]
\begin{center}
\begin{tabular}{||c||c||c||}
\hline\hline
$Quarks$ & $Q_{\psi }$ & $X_{\psi }$ \\ \hline\hline
$%
\begin{tabular}{c}
$q_{L}^{(m)}=\left( 
\begin{array}{c}
u^{(m)} \\ 
d^{(m)} \\ 
J^{(m)}%
\end{array}%
\right) _{L}:3$ \\ 
\\ 
$u_{R}^{(m)},$ $d_{R}^{(m)},$ $J_{R}^{(m)}:1$%
\end{tabular}%
\ \ $ & $%
\begin{tabular}{c}
$\left( 
\begin{array}{c}
\frac{2}{3} \\ 
-\frac{1}{3} \\ 
\frac{2}{3}%
\end{array}%
\right) $ \\ 
\\ 
$\frac{2}{3},-\frac{1}{3},\frac{2}{3}$%
\end{tabular}%
\ \ $ & $%
\begin{tabular}{c}
\\ 
$X_{q^{(m)}}^{L}=\frac{1}{3}$ \\ 
\\ 
\\ 
$X_{q^{(m)}}^{R}=Q_{q^{(m)}}$%
\end{tabular}%
\ \ $ \\ \hline\hline
$%
\begin{tabular}{c}
$q_{L}^{(3^{\ast })}=\left( 
\begin{array}{c}
d^{3^{\ast }} \\ 
-u^{3^{\ast }} \\ 
J^{3^{\ast }}%
\end{array}%
\right) _{L}:3^{\ast }$ \\ 
\\ 
$d_{R}^{3^{\ast }},$ $u_{R}^{3^{\ast }},$ $J_{R}^{3^{\ast }}:1$%
\end{tabular}%
\ \ $ & $%
\begin{tabular}{c}
$\left( 
\begin{array}{c}
-\frac{1}{3} \\ 
\frac{2}{3} \\ 
-\frac{1}{3}%
\end{array}%
\right) $ \\ 
\\ 
$-\frac{1}{3},\frac{2}{3},-\frac{1}{3}$%
\end{tabular}%
\ \ $ & $%
\begin{tabular}{c}
\\ 
$X_{q^{3^{\ast }}}^{L}=0$ \\ 
\\ 
\\ 
$X_{q^{3^{\ast }}}^{R}=Q_{q^{3^{\ast }}}$%
\end{tabular}%
\ \ $ \\ \hline\hline
$%
\begin{tabular}{c}
$q_{L}^{4^{\ast }}=\left( 
\begin{array}{c}
\widetilde{u}^{c} \\ 
\widetilde{d}^{c} \\ 
\widetilde{J}^{c}%
\end{array}%
\right) _{L}:3^{\ast }$ \\ 
\\ 
$\widetilde{u}_{R}^{c},$ $\widetilde{d}_{R}^{c},$ $\widetilde{J}_{R}^{c}:1$%
\end{tabular}%
\ \ $ & $%
\begin{tabular}{c}
$\left( 
\begin{array}{c}
-\frac{2}{3} \\ 
\frac{1}{3} \\ 
-\frac{2}{3}%
\end{array}%
\right) $ \\ 
\\ 
$-\frac{2}{3},\frac{1}{3},-\frac{2}{3}$%
\end{tabular}%
\ \ $ & $%
\begin{tabular}{c}
\\ 
$X_{q^{4^{\ast }}}^{L}=\frac{1}{3}$ \\ 
\\ 
\\ 
$X_{q^{4^{\ast }}}^{R}=Q_{q^{4^{\ast }}}$%
\end{tabular}%
\ \ $ \\ \hline\hline
$Leptons$ & $Q_{\psi }$ & $X_{\psi }$ \\ \hline\hline
\begin{tabular}{c}
$\ell _{L}^{(n)}=\left( 
\begin{array}{c}
\nu ^{(n)} \\ 
e^{(n)} \\ 
N^{0(n)}%
\end{array}%
\right) _{L}:3$ \\ 
\\ 
$\nu _{R}^{(n)},e_{R}^{(n)}:1$%
\end{tabular}
& 
\begin{tabular}{c}
$\left( 
\begin{array}{c}
0 \\ 
-1 \\ 
0%
\end{array}%
\right) $ \\ 
\\ 
$0,-1,0$%
\end{tabular}
& 
\begin{tabular}{c}
\\ 
$X_{\ell ^{(n)}}^{L}=-\frac{1}{3}$ \\ 
\\ 
\\ 
$X_{\ell ^{(n)}}^{R}=Q_{\ell ^{(n)}}$%
\end{tabular}
\\ \hline\hline
\begin{tabular}{c}
$\ell _{L}^{3^{\ast }}=\left( 
\begin{array}{c}
e^{3^{\ast }} \\ 
-\nu ^{3^{\ast }} \\ 
E^{3^{\ast }-}%
\end{array}%
\right) _{L}:3^{\ast }$ \\ 
\\ 
$e_{R}^{3^{\ast }},\nu _{R}^{3^{\ast }},E_{R}^{3^{\ast }-}:1$%
\end{tabular}
& 
\begin{tabular}{c}
$\left( 
\begin{array}{c}
-1 \\ 
0 \\ 
-1%
\end{array}%
\right) $ \\ 
\\ 
$-1,0,-1$%
\end{tabular}
& 
\begin{tabular}{c}
\\ 
$X_{\ell ^{3^{\ast }}}^{L}=\frac{2}{3}$ \\ 
\\ 
\\ 
$X_{\ell ^{3^{\ast }}}^{R}=Q_{\ell ^{3^{\ast }}}$%
\end{tabular}
\\ \hline\hline
\begin{tabular}{c}
$\ell _{L}^{4^{\ast }}=\left( 
\begin{array}{c}
\widetilde{\nu }^{c} \\ 
\widetilde{e}^{c} \\ 
\widetilde{N}^{0c}%
\end{array}%
\right) _{L}:3^{\ast }$ \\ 
\\ 
$\widetilde{\nu }_{R}^{c},\widetilde{e}_{R}^{c}:1$%
\end{tabular}
& 
\begin{tabular}{c}
$\left( 
\begin{array}{c}
0 \\ 
1 \\ 
0%
\end{array}%
\right) $ \\ 
\\ 
$0,1,0$%
\end{tabular}
& 
\begin{tabular}{c}
\\ 
$X_{\ell ^{4^{\ast }}}^{L}=-\frac{1}{3}$ \\ 
\\ 
\\ 
$X_{\ell ^{4^{\ast }}}^{R}=Q_{\ell ^{4^{\ast }}}$%
\end{tabular}
\\ \hline\hline
\end{tabular}%
\end{center}
\caption{\textit{Fermionic content of }$SU\left( 3\right) _{L}\otimes
U\left( 1\right) _{X}$\textit{\ , with }$\mathit{N}=4$, and $\mathit{m,n=1,2}
$\textit{. The 4th families which are in the }$\mathbf{3}^{\ast }\ $\textit{%
representation, are the mirror fermions of one of the families in the }$%
\mathbf{3}$ \textit{representation.}}
\label{fercont4}
\end{table}

We consider a model with $\beta =-1/\sqrt{3}$ which is similar to the model
described in Ref. \cite{twelve} at low energies due to the electromagnetic
charged assigned to different multiplets. However, this model is not the
same as the one in Ref. \cite{twelve} because the multiplets structure for
the quark sector is $SU(3)_{C}\otimes SU(3)_{L}$ vector-like, and the
leptonic part is not neccesary to cancel the quark anomalies. The leptonic
multiplets are also vector-like and anomaly free (see table \ref{fercont4}).
In the models described in the literature, the quarks anomalies are
cancelled out with the leptonic anomalies. In the model with $N=4$ and $%
\beta =-1/\sqrt{3}$ there are two $3$-multiplets for leptons and two $3$%
-multiplets for quarks and they generate the two heavy families of the SM.
Two $3^{\ast }$-multiplets for quarks and leptons correspond to the first SM
family; and the other two $3^{\ast }$, $q_{L}^{4\ast }$ y $l_{L}^{4\ast }$,
correspond to a mirror fermion family of the third SM family. So with this
assignment, it is possible to get mixing between the bottom quark and its
mirror quark $d^{c}$ in order to modify the right-handed coupling of the
bottom quark with the $Z$ gauge boson which in turn might explain the
asymmetry deviations $A^{b}$ and $A_{FB}^{b}$ \cite{Chanowitz}. Such
discrepancy cannot be explained by a model with only left-handed multiplets
such as the SM \cite{anomalia} or the traditional 331 models \cite{Martinez}%
. The mixing in the mass matrix between the $b$ quark and its mirror fermion
permits a solution because the mirror couples with right-handed chirality to
the $Z_{\mu }$ gauge field of the SM. On the other hand, the mirror fermions
in the leptonic sector are useful to build up ansatz about mass matrices in
the neutrino and charged sectors. For the neutrinos corresponding to $%
SU\left( 2\right) _{L}$ doublets, right handed neutrino singlets are
introduced to generate masses of Dirac type.

As for the scalar spectrum, three types of representations are considered.
The three minimal triplets (whose VEV are shown in table \ref{tab:scalaralig}%
) that assure the spontaneous symmetry breaking (SSB)$\ 331\rightarrow
321\rightarrow 31$, and the masses for the gauge fields. Further, an
additional scalar in the adjoint representation is included. Such multiplet
permits a mixing of the mirror fermions with the ordinary fermions of the SM
in order to generate different ansatz for masses. The adjoint representation
acquires the VEV's displayed in table \ref{tab:scalaralig}. Finally, a
sextet representation can also be introduced as shown in table \ref%
{tab:scalaralig}, it acquires very small VEV's compared with the VEV's of
the electroweak scale $\nu _{\chi }$, $\nu _{\rho }$ and $\nu _{\eta }$
since they belong to triplet components of $SU(2)_{L}$ and would not break
the relation for $\Delta \rho $. They also permit to generate majorana
masses for neutrinos.

\begin{table}[tbh]
\begin{center}
\begin{tabular}{||l||l|l||}
\hline\hline
$\left\langle \chi \right\rangle _{0}$ & $\left( 
\begin{array}{ccc}
0 & 0 & \nu _{\chi }%
\end{array}%
\right) ^{T}$ & $X_{\chi }=-1/3$ \\ \hline
$\left\langle \rho \right\rangle _{0}$ & $\left( 
\begin{array}{ccc}
0 & \nu _{\rho } & 0%
\end{array}%
\right) ^{T}$ & $X_{\rho }=2/3$ \\ \hline
$\left\langle \eta \right\rangle _{0}$ & $\left( 
\begin{array}{ccc}
\nu _{\eta } & 0 & 0%
\end{array}%
\right) ^{T}$ & $X_{\eta }=2/3$ \\ \hline\hline
$\left\langle \phi \right\rangle _{0}$ & $\nu _{\chi }diag\left( 
\begin{array}{ccc}
1 & 1 & -2%
\end{array}%
\right) $ & $X_{\chi }=0$ \\ \hline\hline
$\left\langle S^{ij}\right\rangle _{0}$ & $V\left( 
\begin{array}{ccc}
1 & 0 & 0 \\ 
0 & 0 & 0 \\ 
0 & 0 & 1%
\end{array}%
\right) $ & $X_{S}=-1/3$ \\ \hline\hline
\end{tabular}%
\end{center}
\caption{\textit{Scalar sector with }$N=4\ $\textit{and its VEV's. $\protect%
\chi ,\protect\rho ,\protect\eta $ are triplets in the }$\mathit{\mathbf{3}}$
representation, $\protect\phi $ is a multiplet in the adjoint
representation, and $S$ lies in the sextet representation. \textit{$\protect%
\nu _{\protect\chi }$ is of the order of the first symmetry breaking. $%
\protect\nu _{\protect\rho }$, $\protect\nu _{\protect\eta }$ are of the
order of the electroweak scale. $V$ is much lower than the electroweak VEV.}}
\label{tab:scalaralig}
\end{table}

\subsection{Mass matrix for quarks}

The Yukawa Lagrangian for quarks has the form

\begin{eqnarray}
\mathcal{L}_{Y}^{q} &=&\sum_{\Phi }\sum_{sing.}\sum_{m,m^{\prime
}=1}^{2}h_{q_{R}}^{m\varphi }\overline{q_{L}^{(m)}}q_{R}\Phi  \notag \\
&&+\frac{1}{2}\overline{q_{L}^{i(m)}}\left( q_{L}^{j(m^{\prime })}\right)
^{c}\left[ h_{\Phi }^{mm^{\prime }}\varepsilon ^{ijk}\Phi
_{k}+h_{S}^{mm^{\prime }}S^{ij}\right] +h_{q_{R}}^{3\Phi }\overline{%
q_{L}^{(3^{\ast })}}q_{R}\Phi ^{\ast }+h_{q_{R}}^{4\Phi }\overline{%
q_{L}^{(4^{\ast })}}q_{R}\Phi ^{\ast }  \notag \\
&&+\frac{1}{2}\overline{q_{iL}^{(3^{\ast })}}\left( q_{jL}^{(3^{\ast
})}\right) ^{c}\left[ Y_{\Phi }^{33}\varepsilon _{ijk}\Phi
^{k}+Y_{S}^{33}S_{ij}\right] +\frac{1}{2}\overline{q_{iL}^{(4^{\ast })}}%
\left( q_{jL}^{(4^{\ast })}\right) ^{c}\left[ Y_{\Phi }^{44}\varepsilon
_{ijk}\Phi ^{k}+h_{S}^{44}S_{ij}\right]  \notag \\
&&+\frac{1}{2}\overline{q_{iL}^{(3^{\ast })}}\left( q_{jL}^{(4^{\ast
})}\right) ^{c}\left[ Y_{\Phi }^{34}\varepsilon _{ijk}\Phi
^{k}+Y_{S}^{34}S_{ij}\right] +\frac{1}{2}\overline{q_{iL}^{(4^{\ast })}}%
\left( q_{jL}^{(3^{\ast })}\right) ^{c}\left[ Y_{\Phi }^{43}\varepsilon
_{ijk}\Phi ^{k}+Y_{S}^{43}S_{ij}\right]  \notag \\
&&+\frac{1}{2}h_{\phi }^{n3}\overline{q_{L}^{i(n)}}\left( q_{jL}^{(3^{\ast
})}\right) ^{c}\phi _{j}^{i}+\frac{1}{2}h_{\phi }^{3n}\overline{%
q_{iL}^{(3^{\ast })}}\left( q_{L}^{j(n)}\right) ^{c}\phi _{i}^{j}  \notag \\
&&+\frac{1}{2}h_{\phi }^{n4}\overline{q_{L}^{i(n)}}\left( q_{jL}^{(4^{\ast
})}\right) ^{c}\phi _{j}^{i}+\frac{1}{2}h_{\phi }^{4n}\overline{%
q_{iL}^{(4^{\ast })}}\left( q_{L}^{j(n)}\right) ^{c}\phi _{i}^{j}\text{ }%
+h.c,  \label{yukawa-q}
\end{eqnarray}%
with $\Phi $ being any of the $\eta $, $\rho ,$\ $\chi \ $multiplets, while $%
\phi ,\ $and $S$ correspond to the scalar adjoint and the sextet
representation of $SU(3)_{L}$ respectively. The third and fourth families
are written explicitly, since the fourth one correspond to a mirror fermion.
The constants $h_{\Phi }^{mm^{\prime }}$ and $Y_{\Phi }^{34}$ are
antisymmetric. It should be noted that all possible terms with scalar
triplets, adjoints, and sextets are involved. When we take the VEV's from
table \ref{tab:scalaralig}, the mass matrices are obtained.

For the mixing among up-type quarks in the basis $(u_{3^{\ast }},u_{1},u_{2},%
\widetilde{u},J_{1},J_{2},\widetilde{J})$ we get 
\begin{equation}
M^{up}=\left( 
\begin{array}{cc}
\mathcal{M}_{U} & \mathcal{M}_{JU} \\ 
\mathcal{M}_{UJ} & \mathcal{M}_{J}%
\end{array}%
\right) ,  \label{up-mass}
\end{equation}%
where 
\begin{eqnarray*}
\mathcal{M}_{U} &=&\left( 
\begin{array}{cccc}
\nu _{\rho }h_{u_{3}}^{3\rho }\vspace*{0.2cm} & \nu _{\rho }h_{u_{1}}^{3\rho
} & \nu _{\rho }h_{u_{2}}^{3\rho } & h_{\chi }^{43}\nu _{\chi } \\ 
\nu _{\eta }h_{u_{3}}^{1\eta }\vspace*{0.2cm} & \nu _{\eta }h_{u_{1}}^{1\eta
} & \nu _{\eta }h_{u_{2}}^{1\eta } & h_{\phi }^{14}\nu _{\chi } \\ 
\nu _{\eta }h_{u_{3}}^{2\eta }\vspace*{0.2cm} & \nu _{\eta }h_{u_{1}}^{2\eta
} & \nu _{\eta }h_{u_{2}}^{2\eta } & h_{\phi }^{24}\nu _{\chi } \\ 
0 & 0 & 0 & \nu _{\eta }h_{\widetilde{u}}^{4\eta }%
\end{array}%
\right) , \\
\mathcal{M}_{J} &=&\left( 
\begin{array}{ccc}
\nu _{\chi }h_{J_{1}}^{1\varkappa }\vspace*{0.2cm} & \nu _{\chi
}h_{J_{2}}^{1\varkappa } & -2h_{\phi }^{14}\nu _{\chi } \\ 
\nu _{\chi }h_{J_{1}}^{2\varkappa }\vspace*{0.2cm} & \nu _{\chi
}h_{J_{2}}^{2\varkappa } & -2h_{\phi }^{24}\nu _{\chi } \\ 
0 & 0 & \nu _{\chi }h_{\widetilde{J}}^{4\varkappa }%
\end{array}%
\right) ,
\end{eqnarray*}%
\begin{eqnarray*}
\mathcal{M}_{UJ} &=&\left( 
\begin{array}{cccc}
\nu _{\chi }h_{u_{3}}^{1\varkappa }\vspace*{0.2cm} & \nu _{\chi
}h_{u_{1}}^{1\varkappa } & \nu _{\chi }h_{u_{2}}^{1\varkappa } & 0 \\ 
\nu _{\chi }h_{u_{3}}^{2\varkappa }\vspace*{0.2cm} & \nu _{\chi
}h_{u_{1}}^{2\varkappa } & \nu _{\chi }h_{u_{2}}^{2\varkappa } & 0 \\ 
0 & 0 & 0 & \nu _{\eta }h_{\widetilde{J}}^{4\eta }%
\end{array}%
\right) , \\
\mathcal{M}_{JU} &=&\left( 
\begin{array}{ccc}
\nu _{\rho }h_{J_{1}}^{3\rho }\vspace*{0.2cm} & \nu _{\rho }h_{J_{2}}^{3\rho
} & h_{\eta }^{34}\nu _{\eta _{1}} \\ 
\nu _{\eta _{1}}h_{J_{1}}^{1\eta }\vspace*{0.2cm} & \nu _{\eta
_{1}}h_{J_{2}}^{1\eta } & 0 \\ 
\nu _{\eta _{1}}h_{J_{1}}^{2\eta }\vspace*{0.2cm} & \nu _{\eta
_{1}}h_{J_{2}}^{2\eta } & 0 \\ 
0 & 0 & \nu _{\chi }h_{\widetilde{u}}^{4\chi }%
\end{array}%
\right)
\end{eqnarray*}%
and $\left( u_{3^{\ast }},u_{1},u_{2}\right) $ correspond to the three
families of the SM, $\widetilde{u}$ refers to the mirror fermion of either $%
u_{1}$ or $u_{2}$, and $J_{1},J_{2},\widetilde{J}$ are the exotic quarks
with 2/3 electromagnetic charge.

For down-type quarks in the basis $(d_{3^{\ast }},d_{1},d_{2},\widetilde{d}%
,J_{3^{\ast }})$, the mass matrix yields 
\begin{equation}
M^{down}=\left( 
\begin{array}{ccccc}
\nu _{\eta }h_{d_{3}}^{3\eta }\vspace*{0.2cm} & \nu _{\eta }h_{d_{1}}^{3\eta
} & \nu _{\eta }h_{d_{2}}^{3\eta } & Y_{\chi }^{34}\nu _{\chi } & \nu _{\eta
}h_{J_{3}}^{3\eta } \\ 
\nu _{\rho }h_{d_{3}}^{1\rho }\vspace*{0.2cm} & \nu _{\rho }h_{d_{1}}^{1\rho
} & \nu _{\rho }h_{d_{2}}^{1\rho } & h_{\phi }^{14}\nu _{\chi } & \nu _{\rho
}h_{J_{3}}^{1\rho } \\ 
\nu _{\rho }h_{d_{3}}^{2\rho }\vspace*{0.2cm} & \nu _{\rho }h_{d_{1}}^{2\rho
} & \nu _{\rho }h_{d_{2}}^{2\rho } & h_{\phi }^{24}\nu _{\chi } & \nu _{\rho
}h_{J_{3}}^{2\rho } \\ 
0\vspace*{0.2cm} & 0 & 0 & \nu _{\rho }Y_{\widetilde{d}}^{4\rho } & 0 \\ 
\nu _{\chi }h_{d_{3}}^{3\varkappa } & \nu _{\chi }h_{d_{1}}^{3\varkappa } & 
\nu _{\chi }h_{d_{2}}^{3\varkappa } & Y_{\eta }^{43}\nu _{\eta } & \nu
_{\chi }h_{J_{3}}^{3\varkappa }%
\end{array}%
\right)  \label{down-mass}
\end{equation}%
$\left( d_{3^{\ast }},d_{1},d_{2}\right) $ are associated with the three SM
families, $\widetilde{d}$ is a down-type mirror quark of either $d_{1}$or $%
d_{2}$, and $J_{3^{\ast }}$ is an exotic down-type quark. When the adjoint
representation of the scalar fields is not taken into account, the mixing
between $q^{(m)}$ and the quark mirrors $q^{(4^{\ast })}$ does not appear.
Such mixing is important to change the right-handed coupling of the $b-$%
quark with the $Z_{\mu }$ gauge field, and look for a possible solution for
the deviation of the assymmetries $A_{b}$ and $A_{FB}^{b}$ of the SM with
respect to the experimental data. If the mixing with the mirror quarks were
withdrawn and the exotic particles were decoupled, the mirror quarks would
acquire masses of the order of the electroweak scale $\nu _{\rho }h_{%
\widetilde{d}}^{4\rho }$, $\nu _{\eta }h_{\widetilde{u}}^{4\rho }$ for the
up and down sectors, respectively.

\subsection{Mass matrix for Leptons}

The Yukawa Lagrangian for leptons keeps the general form shown in Eq. (\ref%
{yukawa-q}) for the quarks. However, majorana terms could arise because of
the existence of neutral fields. By taking the whole spectrum including
right-handed neutrino singlets, Dirac terms are obtained for the charged
sector while Dirac and majorana terms appear in the neutral sector.

By including all the possible structures of VEV's, the charged sector in the
basis $(e_{3^{\ast }},e_{1},e_{2},\widetilde{e},E_{3^{\ast }}^{-})$ has the
following form

\begin{equation*}
M^{\ell \pm }=\left( 
\begin{array}{ccccc}
\nu _{\eta }h_{e_{3}}^{3\eta }\vspace*{0.2cm} & \nu _{\eta }h_{e_{1}}^{3\eta
} & \nu _{\eta }h_{e_{2}}^{3\eta } & h_{\chi }^{34}\nu _{\chi } & \nu _{\eta
}h_{J_{3}}^{3\eta } \\ 
\nu _{\rho }h_{e_{3}}^{1\rho }\vspace*{0.2cm} & \nu _{\rho }h_{e_{1}}^{1\rho
} & \nu _{\rho }h_{e_{2}}^{1\rho } & h_{\phi }^{14}\nu _{\chi } & \nu _{\rho
}h_{J_{3}}^{1\rho } \\ 
\nu _{\rho }h_{e_{3}}^{2\rho }\vspace*{0.2cm} & \nu _{\rho }h_{e_{1}}^{2\rho
} & \nu _{\rho }h_{e_{2}}^{2\rho } & h_{\phi }^{24}\nu _{\chi } & \nu _{\rho
}h_{J_{3}}^{2\rho } \\ 
0\vspace*{0.2cm} & 0 & 0 & \nu _{\rho }h_{\widetilde{e}}^{4\rho } & 0 \\ 
\nu _{\chi }h_{e_{3}}^{3\varkappa } & \nu _{\chi }h_{e_{1}}^{3\varkappa } & 
\nu _{\chi }h_{e_{2}}^{3\varkappa } & h_{\eta }^{43}\nu _{\eta _{1}} & \nu
_{\chi }h_{E_{3}}^{3\varkappa }%
\end{array}%
\right)
\end{equation*}%
the three first components $e_{i}\ $correspond to the ordinary leptons of
the SM, $\widetilde{e}$ is a mirror lepton of $e_{1}$ or $e_{2}$, and $%
E_{3^{\ast }}$ is an exotic lepton. Like in the case of the quark sector,
direct mixings are gotten between all the fields$\ \ell ^{(n)},\ell
^{3^{\ast }}$ and the mirrors $\ell ^{4^{\ast }}$ by means of the scalars $%
\chi ,\rho ,\eta $ and the adjoint $\phi $. The mass matrix of charged
leptons is similar to the mass matrix of the down-type quarks.

For the neutral lepton sector, we take the following basis of fields

\begin{eqnarray}
\psi _{L}^{0} &=&\left( \nu _{3L},\nu _{1L},\nu _{2L},\left( \widetilde{\nu }%
_{R}\right) ^{c},N_{1L}^{0},N_{2L}^{0},\left( \widetilde{N}_{R}^{0}\right)
^{c}\right) ^{T},  \notag \\
\psi _{R}^{0} &=&\left( \nu _{3R},\nu _{1R},\nu _{2R},\left( \widetilde{\nu }%
_{L}\right) ^{c}\right) ^{T},
\end{eqnarray}%
where $\nu _{iL}$ are the SM fields, $\nu _{iR}$ are sterile neutrinos and
the right handed components of SM neutrinos. With these components the Dirac
mass matrix is constructed like the up quarks mass matrix; $\left( 
\widetilde{\nu }_{L,R}\right) ^{c}$ are mirror fermions, and $N_{iL}^{0}$
are exotic neutral fermions. The mass terms are written as%
\begin{equation}
\mathfrak{L}_{Y}^{0}=\left( \overline{\psi _{L}^{0}};\overline{\left( \psi
_{R}^{0}\right) ^{c}}\right) \left( 
\begin{tabular}{ll}
$M_{L}$ & $m_{D}$ \\ 
$m_{D}^{T}$ & $M_{R}$%
\end{tabular}%
\ \ \right) \left( 
\begin{array}{c}
\left( \psi _{L}^{0}\right) ^{c} \\ 
\psi _{R}^{0}%
\end{array}%
\right) +h.c,  \label{lepton-mass}
\end{equation}%
where very massive majorana terms $M_{R}$ have been introduced between the
singlets $\psi _{R}^{0}$, corresponding to sterile neutrinos with
right-handed chirality. We shall suppose that in this basis the mass matrix $%
M_{R}$ is diagonal. Such terms can be introduced without a SSB because they
are $SU(3)_{L}\otimes U(1)_{X}$ invariant. Besides, they correspond to heavy
majorana mass terms for the sterile heavy neutrinos. The majorana
contribution $M_{L}$ takes the form 
\begin{equation}
M_{L}=\frac{1}{2}\left( 
\begin{array}{cc}
\mathcal{M}_{\nu } & \mathcal{M}_{N\nu } \\ 
\mathcal{M}_{\nu N} & \mathcal{M}_{N}%
\end{array}%
\right) ,
\end{equation}%
where 
\begin{equation*}
\mathcal{M}_{\nu }=\left( 
\begin{array}{cccc}
0\vspace*{0.2cm} & 0 & 0 & -h_{\chi }^{34}\nu _{\chi } \\ 
0\vspace*{0.2cm} & Vh_{S}^{11} & Vh_{S}^{12} & h_{\phi }^{14}\nu _{\chi } \\ 
0\vspace*{0.2cm} & Vh_{S}^{21} & Vh_{S}^{22} & h_{\phi }^{24}\nu _{\chi } \\ 
h_{\chi }^{43}\nu _{\chi } & h_{\phi }^{14}\nu _{\chi } & h_{\phi }^{24}\nu
_{\chi } & Vh_{S}^{44}%
\end{array}%
\right) .
\end{equation*}%
The entries of the upper $3\times 3\ $submatrix correspond to majorana
masses for the ordinary neutrinos of the three SM families, which are
generated with the six dimensional representation of the scalar sector. If
such VEV were taken as null, or if we chose discrete symmetries to forbid
these terms, they can be generated through the see-saw mechanism of the form 
$m_{D}^{\dagger }M_{R}^{-1}m_{D}$. The other mass matrices are given by 
\begin{equation*}
\mathcal{M}_{N}=\left( 
\begin{array}{ccc}
Vh_{S}^{11}\vspace*{0.2cm} & Vh_{S}^{12} & -2h_{\phi }^{14}\nu _{\chi } \\ 
Vh_{S}^{21}\vspace*{0.2cm} & Vh_{S}^{22} & -2h_{\phi }^{24}\nu _{\chi } \\ 
-2h_{\phi }^{14}\nu _{\chi } & -2h_{\phi }^{24}\nu _{\chi } & Vh_{S}^{44}%
\end{array}%
\right) \ ,
\end{equation*}

\begin{equation*}
\mathcal{M}_{\nu N}=\left( 
\begin{array}{cccc}
0\vspace*{0.2cm} & h_{\rho }^{11}\nu _{\rho } & h_{\rho }^{12}\nu _{\rho } & 
0 \\ 
0\vspace*{0.2cm} & h_{\rho }^{21}\nu _{\rho } & h_{\rho }^{22}\nu _{\rho } & 
0 \\ 
-\nu _{\eta _{1}}h_{\eta }^{43} & 0 & 0 & h_{\rho }^{44}\nu _{\rho }%
\end{array}%
\right) \ ,
\end{equation*}

\begin{equation*}
\mathcal{M}_{N\nu }=\left( 
\begin{array}{ccc}
0\vspace*{0.2cm} & 0 & h_{\eta }^{34}\nu _{\eta _{1}} \\ 
-h_{\rho }^{11}\nu _{\rho }\vspace*{0.2cm} & -h_{\rho }^{12}\nu _{\rho } & 0
\\ 
-h_{\rho }^{21}\nu _{\rho }\vspace*{0.2cm} & -h_{\rho }^{22}\nu _{\rho } & 0
\\ 
0 & 0 & -h_{\rho }^{44}\nu _{\rho }%
\end{array}%
\right) .
\end{equation*}%
where we have taken into account the VEV's of the scalar triplets $\chi
,\rho ,\eta $, the adjoint $\phi $ and the sextext $S$. The adjoint VEV's
ensure the direct mixings between $\ell ^{(n)}$ and the mirrors $\ell
^{(4^{\ast })}$. The Dirac terms of (\ref{lepton-mass}) are

\begin{equation}
m_{D}=\frac{1}{2}\left( 
\begin{array}{cccc}
\nu _{\rho }h_{\nu _{3}}^{3\rho }\vspace*{0.2cm} & \nu _{\rho }h_{\nu
_{1}}^{3\rho } & \nu _{\rho }h_{\nu _{2}}^{3\rho } & \nu _{\rho }h_{%
\widetilde{\nu }}^{3\rho } \\ 
\nu _{\eta }h_{\nu _{3}}^{1\eta }\vspace*{0.2cm} & \nu _{\eta }h_{\nu
_{1}}^{1\eta } & \nu _{\eta }h_{\nu _{2}}^{1\eta } & \nu _{\eta }h_{%
\widetilde{\nu }}^{1\eta } \\ 
\nu _{\eta }h_{\nu _{3}}^{2\eta }\vspace*{0.2cm} & \nu _{\eta }h_{\nu
_{1}}^{2\eta } & \nu _{\eta }h_{\nu _{2}}^{2\eta } & \nu _{\eta }h_{%
\widetilde{\nu }}^{2\eta } \\ 
\nu _{\eta }h_{\nu _{3}}^{4\eta } & \nu _{\eta }h_{\nu _{1}}^{4\eta } & \nu
_{\eta }h_{\nu _{2}}^{4\eta } & \nu _{\eta }h_{\widetilde{\nu }}^{4\eta } \\ 
\nu _{\chi }h_{\nu _{3}}^{1\chi }\vspace*{0.2cm} & \nu _{\chi }h_{\nu
_{1}}^{1\chi } & \nu _{\chi }h_{\nu _{2}}^{1\chi } & \nu _{\chi }h_{%
\widetilde{\nu }}^{1\chi } \\ 
\nu _{\chi }h_{\nu _{3}}^{2\chi }\vspace*{0.2cm} & \nu _{\chi }h_{\nu
_{1}}^{2\chi } & \nu _{\chi }h_{\nu _{2}}^{2\chi } & \nu _{\chi }h_{%
\widetilde{\nu }}^{2\chi } \\ 
\nu _{\chi }h_{\nu _{3}}^{4\chi } & \nu _{\chi }h_{\nu _{1}}^{4\chi } & \nu
_{\chi }h_{\nu _{2}}^{4\chi } & \nu _{\chi }h_{\widetilde{\nu }}^{4\chi }%
\end{array}%
\right) .  \label{mD}
\end{equation}

When the quarks and leptons spectra are compared (see table \ref{fercont4}),
it is observed that they are equivalent in the sense that both introduce the
same quantity of particles in the form of left-handed triplets and right
handed singlets (singlet components of neutrinos are taken). Nevertheless,
the Yukawa Lagrangian (and hence the mass matrices) of quarks and leptons
are not equivalent because the quarks have different values of the $X$
quantum number with respect to the leptons, this fact puts different
restrictions on the terms of both Yukawa Lagrangians.

In the limit $\nu _{\rho },\nu _{\eta }<<\nu _{\chi }$ and $V=0$, the
Physics beyond the SM could be decoupled at low energies leaving an
effective theory at low energies similar to a two Higgs doublet model (2HDM)
with the fermionic fields of the SM and the right-handed neutrinos that we
introduced in the particle content $\nu _{1R},\nu _{2R},\nu _{3R}$ to
generate Dirac type masses and be able to relate the neutrino sector with
the up quark sector. It allows to give a large mass to the up quark sector
and the mass pattern for the neutrinos. In this limit, the mass matrices
that are generated would be similar to the ansatz proposed in Ref. \cite%
{soniatwood}. Considering the upper $3\times 3$ submatrix of $m_{D}$ in Eq. (%
\ref{mD}) and imposing discrete symmetries, it can be written in the form 
\begin{equation}
m_{D}=\frac{1}{2}\left( 
\begin{array}{ccc}
\nu _{\rho }h_{\nu _{3}}^{3\rho }\vspace*{0.2cm} & \nu _{\rho }h_{\nu
_{1}}^{3\rho } & 0 \\ 
\nu _{\eta }h_{\nu _{3}}^{1\eta }\vspace*{0.2cm} & \nu _{\eta }h_{\nu
_{1}}^{1\eta } & \nu _{\eta }h_{\nu _{2}}^{1\eta } \\ 
0 & \nu _{\eta }h_{\nu _{1}}^{2\eta } & \nu _{\eta }h_{\nu _{2}}^{2\eta }%
\end{array}%
\right) .  \label{Dirac}
\end{equation}%
considering the same Yukawa couplings within each generation (i.e. the same $%
h_{\nu _{m}}^{n\Phi }$ for each pair $n\Phi $), we can write the matrix (\ref%
{Dirac}) as

\begin{equation}
m_{D}=\frac{\nu _{\eta }}{\sqrt{2}}\left( 
\begin{array}{ccc}
ct_{\beta }\vspace*{0.2cm} & ct_{\beta } & 0 \\ 
\delta b\vspace*{0.2cm} & b & b \\ 
0 & a & a%
\end{array}%
\right) ,
\end{equation}%
where $t_{\beta }=\frac{\nu _{\rho }}{\nu _{\eta }}$ is the scalar mixing
angle given by (\ref{scalar-mixing}), and $\delta $ is a real parameter that
is fitted in agreement with the neutrino oscillation data. If the third
generation is $\nu _{3},$ the second is $\nu _{1}$ and the first is $\nu
_{2},$ and taking $M_{R}=M_{diag}(\epsilon _{M3},\epsilon _{M2},\epsilon
_{M1}),$ we obtain the same mass ansatz and mixing as the Ref. \cite%
{soniatwood}. Thus, from the see-saw mechanism we get

\begin{equation}
m_{\nu }=-m_{D}^{\dagger }M_{R}^{-1}m_{D}=m_{\nu }^{0}\left( 
\begin{array}{ccc}
\delta ^{2}\overline{\epsilon }+\omega \vspace*{0.2cm} & \delta \overline{%
\epsilon }+\omega & \delta \overline{\epsilon } \\ 
\delta \overline{\epsilon }+\omega \vspace*{0.2cm} & \epsilon +\omega & 
\epsilon \\ 
\delta \overline{\epsilon } & \epsilon & \epsilon%
\end{array}%
\right) ,
\end{equation}%
with $m_{\nu }^{0}=\frac{\nu _{\eta }^{2}}{2M},$ $\epsilon =\frac{a^{2}}{%
\epsilon _{M_{1}}}+\frac{b^{2}}{\epsilon _{M_{2}}},$ $\overline{\epsilon }=%
\frac{b^{2}}{\epsilon _{M_{2}}},$ $\omega =\frac{c^{2}t_{\beta }^{2}}{%
\epsilon _{M_{3}}},$ $\tan 2\theta _{23}\sim \frac{2r\omega }{\epsilon
\left( \delta ^{2}-r\right) },$ $\tan 2\theta _{12}\sim \frac{2g}{f},$ $%
\theta _{13}\sim \frac{\epsilon \left( \delta +r\right) }{2^{3/2}r\omega },$ 
$m_{1}\sim \epsilon m_{\nu }^{0}\left\{ 1-g\sin 2\theta _{12}+f\sin
^{2}\theta _{12}\right\} $ , $m_{2}\sim \epsilon m_{\nu }^{0}\left\{ 1+g\sin
2\theta _{12}+f\cos ^{2}\theta _{12}\right\} $, $m_{3}\sim 2\omega m_{\nu
}^{0},$ $r=\frac{\epsilon }{\overline{\epsilon }},$ $g=\frac{\left\vert
r-\delta \right\vert }{\sqrt{2}r},$ and $f=\frac{\delta ^{2}-2\delta -r}{2r}%
. $ As it is discussed in Ref. \cite{soniatwood}, if $m_{3}\sim \sqrt{\Delta
m_{atm}^{2}}$, $m_{2}\sim \sqrt{\Delta m_{sol}^{2}},$ and taking $t_{\beta }=%
\frac{\nu _{\rho }}{\nu _{\eta }}\gg O(1),$ it is possible to obtain a
natural fit for the observed neutrino hierarchical masses and mixing angles.
This result shows the good behavior of the model.

\subsection{The mixing between the bottom quark and its mirror}

In order to look for a solution to the deviation from the $b$ asymmetries,
let us assume that the exotic quarks with charge $1/3$ acquire their mass in
the first SSB and that they are basically decoupled at electroweak energies.
On the other hand, let us suppose that the mass matrix of the three
generations of down quarks is approximately diagonal. In this way the mixing
between the down quark of the third generation ($b$ quark) and its
corresponding mirror can be written as (see Eq. \ref{down-mass})%
\begin{eqnarray}
&&\left( 
\begin{array}{cc}
\bar{d}_{2} & \overline{\widetilde{d}}%
\end{array}%
\right) _{L}\ M\ \left( 
\begin{array}{c}
d_{2} \\ 
\widetilde{d}%
\end{array}%
\right) _{R}\ \ ,  \notag \\
M &\equiv &\left( 
\begin{array}{cc}
h_{d_{2}}^{2\rho }\nu _{\rho } & h_{\phi }^{24}\nu _{\chi } \\ 
0 & Y_{\widetilde{d}}^{4\rho }\nu _{\rho }%
\end{array}%
\right)  \label{defM}
\end{eqnarray}%
The eigenvalues of this mass matrix $M$, that correspond to the masses of
the $b$-quark and the mirror fermion are $h_{d_{2}}^{2\rho }\nu _{\rho }$
and $Y_{\widetilde{d}}^{4\rho }\nu _{\rho }$, respectively. To diagonalize
the mass matrix the following rotation is proposed%
\begin{equation}
\left( 
\begin{array}{c}
b \\ 
\widetilde{b}%
\end{array}%
\right) _{L(R)}=V_{L(R)}^{\dagger }\left( 
\begin{array}{c}
d_{2} \\ 
\widetilde{d}%
\end{array}%
\right) _{L(R)}
\end{equation}%
where $b\ $and $\widetilde{b}$ are the mass eigenstates for the bottom quark
and its mirror fermion respectively. $V_{L}$ and $V_{R}$ are $2\times 2$
matrices of rotation obtained from the matrices $MM^{\dagger }$ and $%
M^{\dagger }M$, respectively (see Eq. \ref{defM}). We shall assume that the
rotation angle of the left-handed quarks ($\theta _{L}$) is small enough,
since it would be tightly restricted by the electroweak processes. For the
right-handed angle we get%
\begin{equation}
\tan 2\theta _{R}=\frac{2h_{\phi }^{24}\nu _{\chi }Y_{\widetilde{d}}^{4\rho
}\nu _{\rho }}{(Y_{\widetilde{d}}^{4\rho }\nu _{\rho
})^{2}-(h_{d_{2}}^{2\rho }\nu _{\rho })^{2}-(h_{\phi }^{24}\nu _{\chi })^{2}}%
\approx \frac{2M_{Z^{\prime }}M_{F}}{M_{F}^{2}-M_{Z^{\prime }}^{2}}
\label{mezclar}
\end{equation}%
in the last line the $b$ quark mass was neglected and the VEV $\nu _{\chi }$
was approximated to$\ M_{Z^{\prime }}$.

When writing the neutral currents for the $d_{2}$ and its mirror $\widetilde{%
d}$ we get 
\begin{eqnarray}
\mathcal{L}_{b}^{NC} &=&\frac{g}{2C_{W}}\overline{d_{2}}\gamma _{\mu }\left[
\left( 1-\frac{2}{3}S_{W}^{2}\right) P_{L}-\frac{2}{3}S_{W}^{2}P_{R}\right]
Z^{\mu }d_{2}  \notag \\
&+&\frac{g}{2C_{W}}\overline{\widetilde{d}}\gamma _{\mu }\left[ \left( 1-%
\frac{2}{3}S_{W}^{2}\right) P_{R}-\frac{2}{3}S_{W}^{2}P_{L}\right] Z^{\mu }%
\widetilde{d}
\end{eqnarray}%
After making the rotations for left and right-handed components of $d_{2}$,$%
\ \widetilde{d}$ quarks, and taking $\theta _{L}=0$, we can write the
right-handed current of the quark bottom mass eigenvalues as 
\begin{equation}
\frac{g}{2C_{W}}\overline{b}\gamma _{\mu }\left( \sin ^{2}\theta _{R}-\frac{2%
}{3}S_{W}^{2}\right) P_{R}Z^{\mu }b
\end{equation}%
and the electroweak right-handed coupling is modified by a factor 
\begin{equation}
\delta g_{R}=\sin ^{2}\theta _{R}
\end{equation}%
By making a combined fit for the LEP and SLD measurements in terms of the
left and right currents of the $b$ quark, and substracting the central value
of the SM it is obtained that\ \cite{valencia2} 
\begin{equation}
\delta g_{R}=0.02
\end{equation}

It means that in order to solve the problem of the deviation of the anomaly $%
A_{b}$, it is necessary for the right-handed mixing angle to be of the order
of $\sin \theta _{R}\approx 0.1$. Replacing this value into Eq. (\ref%
{mezclar}) we find that $M_{Z^{\prime }}\approx 10M_{F}$. This is a
reasonable value if the mirror fermions lie at the electroweak scale and the
first breaking of the 331 model is of the order of the TeV scale.

\section{Conclusions\label{conclusions}}

We have studied the fermionic spectrum of the 331 models with $\beta $
arbitrary by the criterion of cancellation of anomalies. In order to
minimize the exotic spectrum we assume that only one lepton and only one
quark $SU\left( 3\right) _{L}$ multiplet is associated with each generation,
and that there is no more than one right-handed singlet associated with each
left-handed fermion field. By considering models with an arbitrary number of
lepton and quark generations we find the constraints that cancellation of
anomalies provides for the possible fermionic structures. After assuming
that the fermionic $SU\left( 3\right) _{c}\ $representations are
vector-like, and that the SM fermion representations must be embeded in the
triplet 331 representations; we obtain five conditions from the vanishing of
anomalies. The first condition becomes trivial when the SM is embedded in
the 331 model. Two of them restrict the structure of the left-handed
fermionic multiplets, while the other two restrict the structure of
right-handed charged leptonic singlets. The right handed neutral leptonic
singlets are left unconstrained by the equations of anomalies. Under the
assumptions made above, the number of \ left-handed quark multiplets must be
three times the number of left-handed leptonic multiplets because of the
color factor. Besides, models with only one lepton multiplet are forbidden.
In addition, the Higgs and vector spectra, as well as the Yang-Mills
Lagrangian are calculated for $\beta $ arbitrary.

The interest for studying the case of $\beta $ arbitrary is twofold: On one
hand, it permits a general phenomenological analysis that could lead to the
cases studied in the literature. On the other hand, it also permits the
study of other scenarios that could be the source for solving some of the
problems of the SM.

In particular, we studied models with three and four lepton multiplets ($%
N=3,4$). Models with $N=3$ are allowed even if no right-handed charged
leptonic singlets are introduced (Models of Pleitez and Frampton i.e. $\beta
=\pm \sqrt{3}$) as it is indicated in table \ref{tab:N3sol}. However, for
arbitrary values of $\beta $, the three family versions require the
introduction of right-handed charged leptonic singlets in order to cancel
anomalies, and only two type of solutions are possible (see table \ref%
{tab:N3sol}).

The version with $N=4$ and $\beta =-1/\sqrt{3}$, is a vector-like model
consisting of 3 triplets containing the SM fermions plus one triplet
containing mirror fermions of one of the SM families. We choose the mirror
fermions to be associated with the third family of the SM. This $N=4\ $model
is different from similar 331 versions considered in the literature, and
posseses strong phenomenological motivations: the right-handed coupling of
the $b-$quark with the $Z_{\mu }\ $gauge boson could be modified and may in
turn explain the deviation of the $b$ asymmetries with respect to the SM
prediction. In order to solve the $A_{b}$ puzzle, the right-handed mixing
angle should be of the order of $\sin \theta _{R}\approx 0.1$, which in turn
leads to $M_{Z^{\prime }}\approx 10M_{F}$ with $M_{Z^{\prime }}$ and $M_{F}$
denoting the masses for the exotic neutral gauge boson and the mirror
fermion respectively, this relation is reasonable if $M_{F}$ lies in the
electroweak scale and the breaking of the 331 model lies at the TeV scale.
On the other hand, vector-like models are necessary to explain the family
hierarchy. From the phenomenological point of view, the model provides the
possibility of generating ansatz for masses at low energies in the quark and
lepton sector. It worths saying that the Physics beyond the SM could be
decoupled at low energies leaving an effective theory of two Higgs doublets
with right-handed neutrinos, and that the mass matrices generated are
similar to the ansatz proposed by Ref. \cite{soniatwood}. From such ansatz,
a natural fit for the neutrino hierarchical masses and mixing angles can be
obtained.

Finally, this general approach opens a window to analyze other possible 331
versions. For instance, we can analyze the model with $N=4$ but with the
mirror fermion associated with another SM family. Moreover, several models
with $N\geq 4$, with more mirror fermions could be studied from
phenomenological grounds (see table \ref{tab:jkrepres}). In particular, we
observe from table \ref{tab:jkrepres} that $N=6$ contains models that are
vector-like with respect to $SU\left( 3\right) _{L}$ in the quark and lepton
sectors.

\section*{Acknowledgments}

The authors acknowledge to Colciencias and Banco de la Rep\'{u}blica, for
the financial support. R. Martinez also acknowledge the kind hospitality of
Fermilab where part of this work was done.

\appendix

\section{Scalar masses with $\protect\beta $ arbitrary\label{ap:beta}}

\subsection{Imaginary Sector}

The mass matrix is built up in the basis $\zeta _{\chi },\zeta _{\rho
},\zeta _{\eta }:$

\begin{equation}
M_{\zeta \zeta }^{2}=-2f\left[ 
\begin{array}{ccc}
\frac{\nu _{\eta }\nu _{\rho }}{\nu _{\chi }} & \nu _{\eta } & \nu _{\rho }
\\ 
\nu _{\eta } & \frac{\nu _{\eta }\nu _{\chi }}{\nu _{\rho }} & \nu _{\chi }
\\ 
\nu _{\rho } & \nu _{\chi } & \frac{\nu _{\chi }\nu _{\rho }}{\nu _{\eta }}%
\end{array}
\right]  \label{C1}
\end{equation}

The eigenvalues and eigenvectors are given by

\begin{eqnarray}
P_1 &=&P_2=0\ \ ;\ \ P_3=-2f\nu _\chi \left( \frac{\nu _\eta }{\nu _\rho }+%
\frac{\nu _\rho }{\nu _\eta }+\frac{\nu _\eta \nu _\rho }{\nu _\chi ^2}%
\right) ,  \notag \\
\phi _2^0 &=&N_{\phi _2}^0\left( -\nu _\chi \zeta _\chi +\nu _\eta \zeta
_\eta \right) \approx -\zeta _\chi ,  \notag \\
\phi _3^0 &=&N_{\phi _3}^0\left[ -\nu _\chi \nu _\eta ^2\zeta _\chi +\nu
_\rho \left( \nu _\chi ^2+\nu _\eta ^2\right) \zeta _\rho -\nu _\chi ^2\nu
_\eta \zeta _\eta \right] \approx S_\beta \zeta _\rho -C_\beta \zeta _\eta 
\notag \\
h_1^0 &=&C_\beta \zeta _\rho +S_\beta \zeta _\eta ,
\end{eqnarray}
obtained by using the approximations in Eqs. (\ref{hierarchy1}) and (\ref%
{hierarchy2}). The scalars $\phi _2^0$ and $\phi _3^0$ are the would be
Goldstone bosons corresponding to the gauge fields $Z_\mu ^{\prime }$ and $%
Z_\mu $ respectively. $N$ denotes normalizations factors. The mixing angle
is defined by

\begin{equation}
t_{\beta }\equiv \tan \beta =\frac{\nu _{\rho }}{\nu _{\eta }}.
\label{scalar-mixing}
\end{equation}

\subsection{Real sector\label{ap:betare}}

The basis is $\xi _\chi ,\xi _\rho ,\xi _\eta :$ 
\begin{equation}
M_{\xi \xi }^2=\left[ 
\begin{array}{ccc}
8\lambda _1\nu _\chi ^2-2f\frac{\nu _\eta \nu _\rho }{\nu _\chi } & 4\lambda
_4\nu _\chi \nu _\rho +2f\nu _\eta & 4\lambda _5\nu _\chi \nu _\eta +2f\nu
_\rho \\ 
4\lambda _4\nu _\chi \nu _\rho +2f\nu _\eta & 8\lambda _2\nu _\rho ^2-2f%
\frac{\nu _\eta \nu _\chi }{\nu _\rho } & 4\lambda _6\nu _\eta \nu _\rho
+2f\nu _\chi \\ 
4\lambda _5\nu _\chi \nu _\eta +2f\nu _\rho & 4\lambda _6\nu _\eta \nu _\rho
+2f\nu _\chi & 8\lambda _3\nu _\eta ^2-2f\frac{\nu _\chi \nu _\rho }{\nu
_\eta }%
\end{array}
\right]  \label{C2}
\end{equation}
Keeping only quadratic terms in $\nu _\chi $ in the matrix (\ref{C2}), it is
obtained 
\begin{equation}
M_{\xi \xi }^2\simeq \left[ 
\begin{array}{ccc}
8\lambda _1\nu _\chi ^2 & 0 & 0 \\ 
0 & -2f\frac{\nu _\eta \nu _\chi }{\nu _\rho } & 2f\nu _\chi \\ 
0 & 2f\nu _\chi & -2f\frac{\nu _\chi \nu _\rho }{\nu _\eta }%
\end{array}
\right] ,  \label{C8}
\end{equation}
where we get the following decoupled matrices 
\begin{eqnarray}
M_{\xi _\rho \xi _\eta }^2 &\simeq &\left[ 
\begin{array}{cc}
-2f\frac{\nu _\eta \nu _\chi }{\nu _\rho } & 2f\nu _\chi \\ 
2f\nu _\chi & -2f\frac{\nu _\chi \nu _\rho }{\nu _\eta }%
\end{array}
\right] ,  \notag \\
M_{\xi _\chi \xi _\chi }^2 &\rightarrow &8\lambda _1\nu _\chi ^2.  \label{C9}
\end{eqnarray}

The submatrix $M_{\xi _\rho \xi _\eta }^2$ written in Eq. (\ref{C9}) has the
following eigenvalues 
\begin{equation*}
P_2=0\ ;\ P_3=-2f\nu _\chi \left( \frac{\nu _\eta }{\nu _\rho }+\frac{\nu
_\rho }{\nu _\eta }\right) ,
\end{equation*}
The first eigenvalue is zero because of the approximation made in (\ref{C2}%
). If the approximation is not considered, the matrix in (\ref{C9}) takes
the form

\begin{equation}
M_{\xi _{\rho ^{2}}\xi _{\eta }}^{2}=\left[ 
\begin{array}{cc}
8\lambda _{2}\nu _{\rho }^{2}-2f\frac{\nu _{\eta }\nu _{\chi }}{\nu _{\rho }}
& 4\lambda _{6}\nu _{\eta }\nu _{\rho }+2f\nu _{\chi } \\ 
4\lambda _{6}\nu _{\eta }\nu _{\rho }+2f\nu _{\chi } & 8\lambda _{3}\nu
_{\eta }^{2}-2f\frac{\nu _{\chi }\nu _{\rho }}{\nu _{\eta }}%
\end{array}%
\right] ,  \label{C10}
\end{equation}%
and the corresponding eigenvalues are different; they are 
\begin{equation}
P_{2}=\frac{8(\lambda _{2}\nu _{\rho }^{4}+\lambda _{3}\nu _{\eta
}^{4}+\lambda _{6}\nu _{\rho }^{2}\nu _{\eta }^{2})}{\nu _{\eta }^{2}+\nu
_{\rho }^{2}}\ \ ;\ \ P_{3}=-2f\nu _{\chi }\left( \frac{\nu _{\rho }}{\nu
_{\eta }}+\frac{\nu _{\eta }}{\nu _{\rho }}\right) .  \label{eigennonull}
\end{equation}%
and the eigenvectors read 
\begin{equation}
h_{5}^{0}=\xi _{\chi }\ ;\ h_{3}^{0}=S_{\beta }\xi _{\rho }+C_{\beta }\xi
_{\eta }\ ;\ h_{4}^{0}=-C_{\beta }\xi _{\rho }+S_{\beta }\xi _{\eta },
\label{37h}
\end{equation}

\subsection{Charged sector}

The basis is $\chi _{1}^{\pm Q_{1}},\eta _{3}^{\pm Q_{1}}:$

\begin{equation}
M_{\phi \pm Q_1}^2=\left[ 
\begin{array}{cc}
\lambda _7\nu _\eta ^2-f\frac{\nu _\eta \nu _\rho }{\nu _\chi } & \lambda
_7\nu _\chi \nu _\eta -f\nu _\rho \\ 
\lambda _7\nu _\chi \nu _\eta -f\nu _\rho & \lambda _7\nu _\chi ^2-f\frac{%
\nu _\chi \nu _\rho }{\nu _\eta }%
\end{array}
\right] .  \label{C3}
\end{equation}
the mass matrix in the basis $\chi _2^{\pm Q_2},\rho _3^{\pm Q_2}$ is 
\begin{equation}
M_{\phi \pm Q_2}^2=\left[ 
\begin{array}{cc}
\lambda _8\nu _\rho ^2-f\frac{\nu _\eta \nu _\rho }{\nu _\chi } & \lambda
_8\nu _\chi \nu _\rho -f\nu _\eta \\ 
\lambda _8\nu _\chi \nu _\rho -f\nu _\eta & \lambda _8\nu _\chi ^2-f\frac{%
\nu _\chi \nu _\eta }{\nu _\rho }%
\end{array}
\right] .  \label{C4}
\end{equation}
the mass matrix in the basis $\rho _1^{\pm },\eta _2^{\pm }$ reads

\begin{equation}
M_{\phi \pm }^2=\left[ 
\begin{array}{cc}
\lambda _9\nu _\eta ^2-f\frac{\nu _\chi \nu _\eta }{\nu _\rho } & \lambda
_9\nu _\eta \nu _\rho -f\nu _\chi \\ 
\lambda _9\nu _\eta \nu _\rho -f\nu _\chi & \lambda _9\nu _\rho ^2-f\frac{%
\nu _\rho \nu _\chi }{\nu _\eta }%
\end{array}
\right] .  \label{C5}
\end{equation}
it is found that the matrices are singular, it is that 
\begin{equation}
\det (M_{\phi 1\pm }^2)=\det (M_{\phi 2\pm }^2)=\det (M_{\phi 3\pm }^2)=0,
\label{equaldet}
\end{equation}

giving a total of six would be Goldstone bosons. For the matrix $M_{\phi
^{\pm Q_1}}^2$ of Eq. (\ref{C3}) the corresponding eigenvalues and
eigenvectors are found to be

\begin{eqnarray}
P_{1} &=&0\ \ ,\ \ P_{2}=\lambda _{7}\left( \nu _{\eta }^{2}+\nu _{\chi
}^{2}\right) -f\nu _{\rho }\left( \frac{\nu _{\chi }}{\nu _{\eta }}+\frac{%
\nu _{\eta }}{\nu _{\chi }}\right) ,  \notag \\
\phi _{2}^{\pm Q_{1}} &=&N_{\phi _{2}}^{Q_{1}}\left( -\nu _{\chi }\chi
_{1}^{\pm Q_{1}}+\nu _{\eta }\eta _{3}^{\pm Q_{1}}\right) \approx -\chi
_{1}^{\pm Q_{1}},  \notag \\
h_{1}^{\pm Q_{1}} &=&N_{h_{1}}^{\pm Q_{1}}\left( \nu _{\eta }\chi _{1}^{\pm
Q_{1}}+\nu _{\chi }\eta _{3}^{\pm Q_{1}}\right) \approx \eta _{3}^{\pm
Q_{1}},  \label{37b}
\end{eqnarray}%
where the approximations of Eqs. (\ref{hierarchy1}) and (\ref{hierarchy2})
have been taken into account; getting two would-be Goldstone bosons $\phi
_{2}^{\pm Q_{1}}$ associated with the gauge fields $K_{\mu }^{\pm Q_{1}}$
and two massive Higgs bosons $h_{1}^{\pm Q_{1}}$.

For $M_{\phi ^{\pm Q_{2}}}^{2}$, from Eq. (\ref{C4}), we find 
\begin{eqnarray}
P_{3} &=&0\ \ ,\ \ P_{4}=\lambda _{8}\left( \nu _{\rho }^{2}+\nu _{\chi
}^{2}\right) -f\nu _{\eta }\left( \frac{\nu _{\chi }}{\nu _{\rho }}+\frac{%
\nu _{\rho }}{\nu _{\chi }}\right) ,  \notag \\
\phi _{3}^{\pm Q_{2}} &=&N_{\phi _{3}}^{Q_{2}}\left( -\nu _{\chi }\chi
_{2}^{\pm Q_{2}}+\nu _{\rho }\rho _{3}^{\pm Q_{2}}\right) \approx -\chi
_{2}^{\pm Q_{2}},  \notag \\
h_{3}^{\pm Q_{2}} &=&N_{h_{3}}^{Q_{2}}\left( \nu _{\rho }\chi _{2}^{\pm
Q_{2}}+\nu _{\chi }\rho _{3}^{\pm Q_{2}}\right) \approx \rho _{3}^{\pm Q_{2}}
\end{eqnarray}
again we have used the approximations in Eqs. (\ref{hierarchy1}) and (\ref%
{hierarchy2}); obtaining two would-be Goldstone bosons $\phi _{3}^{\pm
Q_{2}} $ associated with the gauge fields $K_{\mu }^{\pm Q_{2}}$ and two
massive Higgs bosons $h_{3}^{\pm Q_{2}}$.

Finally, for $M_{\phi ^{\pm }}^{2}$ and from Eq. (\ref{C5}) we have:

\begin{eqnarray}
P_{5} &=&0\ \ ,\ \ P_{6}=\lambda _{9}\left( \nu _{\rho }^{2}+\nu _{\eta
}^{2}\right) -f\nu _{\chi }\left( \frac{\nu _{\eta }}{\nu _{\rho }}+\frac{%
\nu _{\rho }}{\nu _{\eta }}\right) ,  \notag \\
\phi _{1}^{\pm } &=&S_{\beta }\rho _{1}^{\pm }-C_{\beta }\eta _{2}^{\pm }\ \
,\ \ h_{2}^{\pm }=C_{\beta }\rho _{1}^{\pm }+S_{\beta }\eta _{2}^{\pm },
\label{charged Higgses}
\end{eqnarray}%
where $\phi _{1}^{\pm }$ give mass to $W_{\mu }^{\pm }$. 

\section{The mass matrix for the neutral gauge sector \label%
{massneutralgauge}}

The basis for the mass matrix for the neutral gauge sector is $W^3,W^8,B:$ 
\begin{eqnarray*}
&& \\
&&\hspace{-3cm}\left[ 
\begin{array}{ccc}
\frac{g^2}2\left( \nu _\eta ^2+\nu _\rho ^2\right) & \frac{g^2}{2\sqrt{3}}%
\left( \nu _\eta ^2-\nu _\rho ^2\right) & 
\begin{tabular}{l}
$\frac{gg^{\prime }}2\left[ -\nu _\eta ^2\left( 1+\frac \beta {\sqrt{3}%
}\right) \right. $ \\ 
$\left. -\nu _\rho ^2\left( 1-\frac \beta {\sqrt{3}}\right) \right] $%
\end{tabular}
\\ 
\begin{tabular}{c}
\vspace{0.5cm} \\ 
$\frac{g^2}{2\sqrt{3}}\left( \nu _\eta ^2-\nu _\rho ^2\right) $%
\end{tabular}
& 
\begin{tabular}{c}
\vspace{0.5cm} \\ 
$\frac{g^2}6\left( \nu _\eta ^2+\nu _\rho ^2+4\nu _\chi ^2\right) $%
\end{tabular}
& 
\begin{tabular}{c}
\vspace{0.5cm} \\ 
$\frac{gg^{\prime }}6\left[ -\nu _\eta ^2\left( \sqrt{3}+\beta \right)
\right. $ \\ 
$\left. +\nu _\rho ^2\left( \sqrt{3}-\beta \right) -4\nu _\chi ^2\beta %
\right] $%
\end{tabular}
\\ 
\begin{tabular}{c}
\vspace{0.5cm} \\ 
$\frac{gg^{\prime }}2\left[ -\nu _\eta ^2\left( 1+\frac \beta {\sqrt{3}%
}\right) \right. $ \\ 
$\left. -\nu _\rho ^2\left( 1-\frac \beta {\sqrt{3}}\right) \right] $%
\end{tabular}
& 
\begin{tabular}{c}
\vspace{0.5cm} \\
$\frac{gg^{\prime }}6\left[ -\nu _\eta ^2\left( \sqrt{3}+\beta \right)
\right. $ \\
$\left. +\nu _\rho ^2\left( \sqrt{3}-\beta \right) -4\nu _\chi ^2\beta %
\right] $%
\end{tabular}
&
\begin{tabular}{c}
\vspace{0.5cm} \\
$\frac{g^{\prime 2}}6\left[ \nu _\eta ^2\left( \sqrt{3}+\beta \right)
^2\right. $ \\
$\left. +\nu _\rho ^2\left( \sqrt{3}-\beta \right) ^2+4\nu _\chi ^2\beta ^2%
\right] $%
\end{tabular}%
\end{array}
\right]
\end{eqnarray*}

\end{document}